\documentclass[aps,preprintnumbers,nofootinbib,twocolumn,amsmath,amssymb,floatfix,superscriptaddress,pra]{revtex4-2}
\usepackage[utf8]{inputenc}
\usepackage[T1]{fontenc}
\usepackage[german,english]{babel}
\usepackage{graphicx}
\usepackage{amsmath}
\usepackage{mathtools}
\usepackage{amssymb}
\usepackage{amsthm}
\usepackage{tensor}
\usepackage{microtype} 
\usepackage[caption = false]{subfig}
\usepackage{quoting} 
\quotingsetup{font=small}
\usepackage{tensor}
\usepackage{graphicx}
\usepackage{braket}
\usepackage{bm}
\usepackage{comment}
\usepackage{booktabs}
\usepackage{braket}
\usepackage{bbold} 
\usepackage{emptypage}
\usepackage{epstopdf}
\usepackage{verbatim}
\usepackage{enumitem}
\usepackage{float}
\usepackage{chemformula}
\usepackage{tikz}
\usepackage{pgfplots}

\newcommand{\beq}{\begin{equation}}
\newcommand{\eeq}{\end{equation}}

\pgfplotsset{compat=newest}
\pgfplotsset{
    height=7.5cm, width=7.5cm, tick label style={font=\huge}, label style={font=\LARGE}, legend style={font=\LARGE}, every tick/.style={black,semithick}, xtick distance=10^(3), ytick distance=10^3}

\usetikzlibrary{shapes.geometric}
\usetikzlibrary{decorations.markings}
\usetikzlibrary{positioning ,arrows, external, backgrounds}
\usetikzlibrary{datavisualization.formats.functions} 

\usepgfplotslibrary{patchplots}

\usepackage[bookmarks=true,colorlinks,linkcolor=red,urlcolor=blue,citecolor=blue]{hyperref} 
\usepackage{cleveref}

\begin{document}
\title{Quantum reaction-limited reaction-diffusion dynamics of noninteracting Bose gases}
\author{Shiphrah Rowlands}
\affiliation{Institut f\"ur Theoretische Physik, Universit\"at T\"ubingen, Auf der Morgenstelle 14, 72076 T\"ubingen, Germany}
\author{Igor Lesanovsky}
\affiliation{Institut f\"ur Theoretische Physik, Universit\"at T\"ubingen, Auf der Morgenstelle 14, 72076 T\"ubingen, Germany}
\affiliation{School of Physics and Astronomy and Centre for the Mathematics and Theoretical Physics of Quantum Non-Equilibrium Systems, The University of Nottingham, Nottingham, NG7 2RD, United Kingdom}
\author{Gabriele Perfetto}
\affiliation{Institut f\"ur Theoretische Physik, Universit\"at T\"ubingen, Auf der Morgenstelle 14, 72076 T\"ubingen, Germany}

\begin{abstract}
We investigate quantum reaction-diffusion systems in one-dimension with bosonic particles that coherently hop in a lattice, and when brought in range react dissipatively. Such reactions involve binary annihilation ($A + A \to \emptyset$) and coagulation ($A + A \to A$) of particles at distance $d$. 
We consider the reaction-limited regime, where dissipative reactions take place at a rate that is small compared to that of coherent hopping. In classical reaction-diffusion systems, this regime is correctly captured by the mean-field approximation. In quantum reaction-diffusion systems, for noninteracting fermionic systems, the reaction-limited regime recently attracted considerable attention because it has been shown to give universal power law decay beyond mean field for the density of particles as a function of time. Here, we address the question whether such universal behavior is present also in the case of the noninteracting Bose gas. 
We show that beyond mean-field density decay for bosons is possible only for reactions that allow for destructive interference of different decay channels. 
Furthermore, we study an absorbing-state phase transition induced by the competition between branching $A\to A+A$, decay $A\to \emptyset$ and coagulation $A+A\to A$. We find a stationary phase-diagram, where a first and a second-order transition line meet at a bicritical point which is described by tricritical directed percolation. 
These results show that quantum statistics significantly impact on both the stationary and the dynamical universal behavior of quantum reaction-diffusion systems. 
\end{abstract}
\maketitle

\section{Introduction}
The systematic classification of universal properties in nonequilibrium many-body systems is a timely and challenging research direction in statistical mechanics. In this field, reaction-diffusion (RD) models are an important class of classical many-body nonequilibrium systems where dynamical critical behavior has been thoroughly studied \cite{vladimir1997nonequilibrium, henkel2008non, hinrichsen2000non,krapivsky2010kinetic,tauber2014critical,Malte_Henkel_2004}. These systems describe particles which diffuse at rate $\Omega$ through random motion in space and, when brought in contact, can react with each other at rate $\Gamma$. The particle density $n(t)$ of RD systems at long times $t$ decays as a power law. The latter is described in the ``reaction-limited regime'' of strong diffusion $\Gamma \ll \Omega$ by mean-field approximation \cite{hinrichsen2000non,vladimir1997nonequilibrium,Redner1984, fastdiffusion1992,krapivsky2010kinetic,Urbano2023}. Fast diffusion, indeed, quickly erases spatial structures rendering the density profile homogeneous in space. Under these conditions, the dynamics is described by law of mass action rate equations, which for reactions as $mA\to lA$ ($l<m$) predict
\begin{equation}
n(t)\sim (\Gamma t)^{-1/(m-1)}.
\label{eq:law_of_action_intro}
\end{equation}
In particular, annihilation $2A\to \emptyset$ at rate $\Gamma_{\alpha}$ ($m=2$, $l=1$) and coagulation $2A\to A$ ($m=2$, $l=1$) at rate $\Gamma_{\gamma}$ within mean field behave as $n(t)\sim (\Gamma_{\alpha,\gamma} t)^{-1}$. In the opposite ``diffusion-limited regime'' $\Gamma/\Omega \sim 1$, instead, spatial fluctuations are relevant. Annihilation and coagulation, for instance, belong the same universality class and in one dimension, $D=1$, in both the cases one finds algebraic decay \cite{toussaint1983particle,Spouge1988,Privman1994,torney1983diffusion,fluctuationseffects,redner1984scaling,kang1984universal,Racz1985,tauber2002dynamic,tauber2005applications,tauber2014critical,doi1976stochastic,doi1976,peliti1985path,peliti1986renormalisation,QFT_RD_1998} beyond mean field 
\begin{equation}
n(t)\sim (\Omega t)^{-1/2}.
\label{eq:diffusion_limited_intro}
\end{equation}
For $D>1$, instead, the mean-field prediction \eqref{eq:law_of_action_intro} is retrieved as diffusion is effective in filling the whole available space thus eventually rendering the density homogeneous. Reaction-diffusion systems display also stationary-state universal properties. This happens when branching processes $A\to A+A$, at rate $\Gamma_{\beta}$, injecting particles into the system are considered. Namely, in the contact process (CP) \cite{hinrichsen2000non,henkel2008non,tauber2014critical,Malte_Henkel_2004}, the competition between branching and one-body decay $A\to\emptyset$, at rate $\Gamma_{\delta}$, induces a stationary-state phase transition. In the thermodynamic limit and at long times, tuning the relative strength $\Gamma_{\beta}/\Gamma_{\delta}$ between the branching and the decay rate, one has either an active phase, where the stationary density is nonzero, or an inactive phase with zero density. In the latter phase, the stationary state is an absorbing-state void of particles and no fluctuations are possible. The absorbing-state phase transition of the contact process is of second-order and it belongs to the directed percolation universality class \cite{hinrichsen2000non,vladimir1997nonequilibrium,Redner1984, fastdiffusion1992,krapivsky2010kinetic}. For restricted occupancy of the lattice, the stationary density in the active phase is finite, while in the unrestricted case one needs to include also annihilation and/or coagulation in order to have a finite stationary density \cite{tauber2005applications,Malte_Henkel_2004,Paessens_2004}.    

Quantum RD systems have been only recently been addressed and studied in Refs.~\cite{RDHorssen,lossth4,lossth5,lossth7,lossth8,lossth8bis,lossth9,lossth11,lossth1,lossth2,lossth6,lossth3,QRD20222,perfetto2023quantum,riggio2023effects,gerbino2023large}. The dynamics is formulated in terms of the Markovian quantum master equation \cite{lindblad1976,gorini1976,breuer2002}, where coherent-Hamiltonian hopping replaces stochastic diffusion while reactions are dissipative. These quantum RD models attract significant attention since they connect the physics of Markovian open quantum systems \cite{griessner2006,diehl2008,kraus2008,diehl2011,tomadin2011,bardyn2013,perez-espigares2017,buca2020,carollo2022} to that of systems with kinetic constraints \cite{lesanovsky2013,olmos2014,everest2016,marcuzzi2016,buchhold2017,gutierrez2017,roscher2018,carollo2019,gillman2019,gillman2020,wintermantel2020,helmrich2020,nigmatullin2021,kazemi2021,carollo2022quantum}. From the experimental perspective, on the one hand, quantum RD systems naturally connect to cold-atomic experiments involving particle losses \cite{lossexp0,lossexp3,lossexp4,lossexpF1,lossexpF2,lossexpF3,lossexp5,lossexp5bis}. From the theoretical perspective, on the other hand, quantum RD systems are interesting since they allow to investigate whether and how quantum effects alter the universal properties of the dynamics. However, a fair assessment of out-of-equilibrium universal properties necessarily requires considering simultaneously large system sizes and long times. This regime is hard to treat due to the exponential complexity of many-body quantum simulations. For this reason, quantum RD models provide a natural test-bed for quantum simulators \cite{quantum_simulator_1, quantum_simulator_2, quantum_simulator_3, quantum_simulator_4} and large-scale numerical simulations \cite{RDHorssen,carollo2019,gillman2019,gillman2020,jo2021}. No analytical prediction for the quantum analogue of the diffusion-limited regime \eqref{eq:diffusion_limited_intro} is, indeed, currently available. 

Quantum RD systems have been treated analytically, in the thermodynamic limit, only in the reaction-limited regime in Refs.~\cite{lossth1,lossth2,lossth3,lossth6,lossth11,QRD20222,perfetto2023quantum,gerbino2023large} for fermionic systems both in the lattice \cite{lossth1, lossth6,QRD20222,lossth11,perfetto2023quantum} and in the continuum \cite{lossth2,lossth3,gerbino2023large}. Spin systems subject to $m$-body losses have been also considered in Ref.~\cite{riggio2023effects}. The studies \cite{lossth1,lossth2,lossth3,lossth6,QRD20222,perfetto2023quantum,gerbino2023large,riggio2023effects} are all based on the time-dependent generalized Gibbs ensemble (TGGE) method \cite{tGGE1,tGGE2,tGGE3,tGGE4,tGGE5}, while \cite{gerbino2023large} employs Keldysh field-theory diagrammatics to show the emergence of the TGGE dynamical equation. It has been found in these references that universal decay beyond mean field is present for initial states displaying quantum coherences in real space. The law of mass action prediction \eqref{eq:law_of_action_intro} is, instead, recovered when the dynamics starts from incoherent initial states. This is specifically true for annihilation processes of $m$ neighbouring particles $mA \to \emptyset$ \cite{QRD20222,perfetto2023quantum,riggio2023effects} (on site multibody losses are not possible for fermions). For $2A \to \emptyset$, for coherent initial states, one specifically finds $n(t)\sim (\Gamma_{\alpha}t)^{-1/2}$. Coagulation reactions $2A\to A$, on the contrary, decay as $n(t) \sim (\Gamma_{\gamma}t)^{-1}$ with the mean-field exponent is found both for coherent and incoherent initial states. This, in particular, shows that for quantum systems binary annihilation and coagulation do not belong to the same universality class, in stark contrast to the classical case. In Refs.~\cite{lossth1,QRD20222}, annihilation reactions that create quantum superpositions between two nearest-neighbor decay channels were considered. Annihilation jump operators with this feature introduce inherent quantum interference effects and allows for a non-mean-field decay $n(t)\sim (\Gamma_{\alpha}t)^{-1/2}$ for both coherent and incoherent initial states. For the case where the CP is considered with branching and decay, 
the associated absorbing-state phase transition is of second order and it is found to belong to the mean-field directed percolation universality class. 

The reaction-limited dynamics of the interacting Bose gas has been studied in Ref.~\cite{lossth3}. In the limiting case of no interaction, it is therein found that the density for onsite losses $mA\to\emptyset$ decays according to the law of mass action \eqref{eq:law_of_action_intro}. For other kinds of reactions additional analysis is, however, needed. In particular, it is interesting to understand whether reactions coupling bosons at different sites possibly creating superpositions between two, or more, decay channels induce beyond mean-field decay, as in the aforementioned fermionic cases. Furthermore, the quantum reaction-limited dynamics of the Bose gas in the presence of reactions, such as branching, creating particles has not been analyzed so far. In this perspective, for bosons, the universality class and the order of the absorbing-state phase transition taking place in the presence of branching is also not understood. 

The goal of this manuscript is to address the impact of quantum bosonic statistics on the universal behavior of the quantum RD dynamics. In particular, we study the impact of the bosonic statistics both on the relaxation dynamics of the noninteracting Bose gas and on the universality class of its stationary phase diagram obtained in the presence of branching reactions. In both the relaxation and the stationary case, our goal is then to show that the universal properties of quantum RD dynamics of the Bose gas significantly differ both from the fermionic formulation and from their classical counterpart.

Our approach is based on analytically studying the quantum RD dynamics of the noninteracting Bose gas in the reaction-limited regime and in the thermodynamic limit. We employ the time-dependent generalised Gibbs ensemble (TGGE) method \cite{tGGE1, tGGE2,tGGE3,tGGE4,tGGE5}. We first consider distance-selective binary losses, which amount to a loss of two bosons at a distance $d$. We further allow for the possibility of interference between two or three decay channels of the distance selective loss. We then move to the case of onsite coagulation reactions and eventually consider the absorbing-state phase transition induced by the competition between coagulation, branching and one-body decay. 

We find that distance selective losses eventually always render, independently of the initial state, the law of mass action prediction
\begin{equation}
n(t)\sim (\Gamma_{\alpha} t)^{-1}, \quad \mbox{distance-selective} \,\,\,  2A\to \emptyset,
\label{eq:intro_distance_selective_MF}
\end{equation}
for any value of the distance $d$. This is in contrast to the case of fermionic systems, where non-mean-field behavior is found for nearest neighbor annihilation ($d=1$). The case where interference between two annihilation decay channels is present is then discussed. When the channels are unequally weighted, we retrieve mean-field behaviour as in Eq.~\eqref{eq:intro_distance_selective_MF}. For a fine-tuned equal balance between the two channels, instead, we find algebraic decay beyond mean-field for any value $d \geq 1$. Namely, we observe the same asymptotic decay as in Refs.~\cite{lossth1,QRD20222} for fermions,
\begin{equation}
n(t)\sim (\Gamma_{\alpha}t)^{-1/2}, \,\, \mbox{two-channels interference}, \,\, 2A\to \emptyset.
\label{eq:intro_distance_interf_1}
\end{equation}
The main difference with respect to the fermionic case is that in the latter case Eq.~\eqref{eq:intro_distance_interf_1} does not require the two decay channels to be equally weighted. 
We eventually consider the case where the interference takes place between three neighbouring decay channels (rate $\Gamma_{\Bar{\alpha}}$). In this case, we also find beyond mean-field decay 
\begin{equation}
n(t) \sim (\Gamma_{\Bar{\alpha}} t)^{-0.28}, \,\, \mbox{three-channels interference}, \, 2A\to\emptyset.
\end{equation}
for any $d\geq 1$. These results show that for bosons interference effects within the annihilation decay channels are necessary in order to get beyond mean-field algebraic decays. 
For onsite coagulation, we find mean-field decay 
\begin{equation}
n(t)\sim (\Gamma_{\gamma}t)^{-1},
\label{eq:intro_coagulation}
\end{equation}
when coagulation is considered alone. This decay is valid independently of the initial state considered. 
Equation \eqref{eq:intro_coagulation} implies that for bosons annihilation \eqref{eq:intro_distance_selective_MF} and coagulation share the same decay exponent and they belong, at least in the reaction-limited regime, to the same universality class. This again contrasts the fermionic case of Ref.~\cite{QRD20222}, where the same reaction-limited decay exponent is observed only for incoherent initial states. 

The competition between coagulation, one-body decay and branching in the CP reveals gives rise to an interesting phase diagram. It is of mean-field nature but considerably richer than the phase diagram obtained within the classical and fermionic reaction-limited cases. 
In particular, we find a change of the absorbing-state phase transition from first to second order as the relative strength $\Gamma_{\beta}/\Gamma_{\delta}$ between classical branching and decay is tuned. These first and second-order transition lines meet at a bicritical point, which is characterized by the mean-field exponents of tricritical directed percolation \cite{grassberger1982,lubeck2006tricritical,grassberger2006tricritical}. Mean-field exponents are here observed already in one dimensions since the reaction-limited regime with fast hopping mixing is considered. A similar phase diagram has been previously observed in Refs.~\cite{marcuzzi2016,buchhold2017} for a different spin $1/2$ model, where classical branching/coagulation competes with quantum branching/coagulation. Therein the order of the transition is changed as the relative strength between coherent and incoherent branching is tuned. The phase transitions observed in our work, however, solely rely on classical reactions, with the only quantum effect stemming from the hopping of the bosons.

This manuscript is structured as follows. In Sec.~\ref{sec:system}, we discuss the formulation of quantum RD dynamics in terms of the quantum master equation. We further present all the reactions processes analysed in the text. In Sec.~\ref{sec:rd_dynamics}, we briefly review known results of classical RD dynamics and then we discuss the quantum RD dynamics. In particular, we focus on the reaction-limited regime, which is the main subject of the text. We briefly introduce the TGGE method and we write the associated dynamical equation for the bosonic occupation function in momentum space. In Sec.~\ref{sec:results}, we specialize this equation to the various processes introduced in Sec.~\ref{sec:system}. In Sec.~\ref{sec:discussion}, we summarize and discuss our results. The Appendices \ref{app:cont_limit}-\ref{app:branching_bosons} contain technical aspects and details regarding the calculations at the basis of the results presented in the main text.

\section{The system} \label{sec:system}
\begin{figure}[t]
    \centering
    \resizebox{0.9\columnwidth}{!}{\tikzsetnextfilename{setup}

\tikzstyle{occupied}=[draw, circle, fill=black!40!white,inner sep=1pt]
\tikzstyle{occupied_less}=[draw, circle, fill=black!10!white,inner sep=1pt]
\tikzstyle{occupied_more}=[draw, circle, fill=black!70!white,inner sep=1pt]
\tikzstyle{empty}=[draw, circle, fill=white,inner sep=1pt]

\begin{tikzpicture}
\node[black, thick] at (0,1.04) {\large Hopping};
\node[occupied, scale=3] at (-1,0){};
\node[occupied_more, scale=3] at (-0.6,0){};
\draw[<->, decorate, draw=teal, thick] (-0.4,0) -- (0.4,0) node[midway,above,teal] {\large $\Omega$};
\node[occupied_more, scale=3] at (0.6,0){};
\node[occupied, scale=3] at (1,0){};
    \node[black, thick] at (2.6,1) {\large Coagulation};
    \node[occupied_more, scale=3] at (2,0){};
    \draw[->, decorate, decoration={snake}, draw=blue, thick] (2.2,0) -- (3,0) node[midway,above, blue] {\large $\Gamma_{\gamma}$};
    \node[occupied, scale=3] at (3.2,0){};
    \node[black, thick] at (5,1) {\large Decay};
    \node[occupied, scale=3] at (4.5,0){};
    \draw[->, decorate, decoration={snake}, draw=blue, thick] (4.7,0) -- (5.5,0) node[midway,above, blue] {\large $\Gamma_{\delta}$};
    \node[empty, scale=3] at (5.7,0){};
    \node[black, thick] at (7.3,1) {\large Branching};
    \node[occupied, scale=3] at (6.9,0){};
    \draw[->, decorate, decoration={snake}, draw=blue, thick] (7.1,0) -- (7.9,0) node[midway,above, blue] {\large $\Gamma_{\beta}$};
    \node[occupied_more, scale=3] at (8.1,0){};
    \node[black, thick] at (1.5,-1) {\large Two-channel annihilation};
    \draw[thick] (-0.3,-1.7) -- (-0.3,-2.2) {};
    \node[occupied_more, scale=3] at (-0.1,-1.95) {};
    \node[occupied_more, scale=3] at (0.3,-1.95) {};
    \node[occupied_more, scale=3] at (0.7,-1.95) {};
    \draw[thick] (0.9,-1.7) -- (1,-1.95) {};
    \draw[thick] (1,-1.95) -- (0.9,-2.2) {};
    \draw[->, decorate, decoration={snake}, draw=blue, thick] (1.2,-2) -- (2.3,-2) node[midway,above, blue] {\large $\Gamma_{\alpha}$};
    \draw[thick] (3.7,-1.7) -- (3.7,-2.2) {};
    \node[occupied_more, scale=3] at (3.9,-1.95) {};
    \node[occupied, scale=3] at (4.3,-1.95) {};
    \node[occupied, scale=3] at (4.7,-1.95) {};
    \draw[thick] (4.9,-1.7) -- (5,-1.95) {};
    \draw[thick] (5,-1.95) -- (4.9,-2.2) {};
    \draw[thick] (6.7,-1.7) -- (6.7,-2.2) {};
    \node[occupied, scale=3] at (6.9,-1.95) {};
    \node[occupied, scale=3] at (7.3,-1.95) {};
    \node[occupied_more, scale=3] at (7.7,-1.95) {};
    \draw[thick] (7.9,-1.7) -- (8,-1.95) {};
    \draw[thick] (8,-1.95) -- (7.9,-2.2) {};
    \node[scale=1.1] at (3.1,-1.98) {$\cos(\theta)$};
    \node[scale=1.1] at (6.2,-1.98) {$\sin(\theta)$};
    \node[scale=1.5] at (5.4,-2) {$+$};
    \node[black, thick] at (1.7,-3) {\large Three-channel annihilation};
    \draw[thick] (-0.3,-3.7) -- (-0.3,-4.2) {};
    \node[occupied_more, scale=3] at (-0.1,-3.95) {};
    \node[occupied_more, scale=3] at (0.3,-3.95) {};
    \node[occupied_more, scale=3] at (0.7,-3.95) {};
    \draw[thick] (0.9,-3.7) -- (1,-3.95) {};
    \draw[thick] (1,-3.95) -- (0.9,-4.2) {};
    \draw[->, decorate, decoration={snake}, draw=blue, thick] (1.2,-4) -- (2.3,-4) node[midway,above, blue] {\large $\Gamma_{\Bar{\alpha}}$};
    \draw[thick] (2.6,-3.7) -- (2.6,-4.2) {};
    \node[occupied_more, scale=3] at (2.8,-3.95) {};
    \node[occupied, scale=3] at (3.2,-3.95) {};
    \node[occupied, scale=3] at (3.6,-3.95) {};
    \draw[thick] (3.8,-3.7) -- (3.9,-3.95) {};
    \draw[thick] (3.9,-3.95) -- (3.8,-4.2) {};
    \node[scale=1.5] at (4.2,-3.95) {$-$};
    \node[scale=1.3] at (4.5,-3.94) {$2$};
    \draw[thick] (4.8,-3.7) -- (4.8,-4.2) {};
    \node[occupied_more, scale=3] at (5,-3.95) {};
    \node[occupied_less, scale=3] at (5.4,-3.95) {};
    \node[occupied_more, scale=3] at (5.8,-3.95) {};
    \draw[thick] (6,-3.7) -- (6.1,-3.95) {};
    \draw[thick] (6.1,-3.95) -- (6,-4.2) {};
    \node[scale=1.5] at (6.4,-3.95) {$+$};
    \draw[thick] (6.7,-3.7) -- (6.7,-4.2) {};
    \node[occupied, scale=3] at (6.9,-3.95) {};
    \node[occupied, scale=3] at (7.3,-3.95) {};
    \node[occupied_more, scale=3] at (7.7,-3.95) {};
    \draw[thick] (7.9,-3.7) -- (8,-3.95) {};
    \draw[thick] (8,-3.95) -- (7.9,-4.2) {};
\end{tikzpicture}}
    \caption{\textbf{Quantum RD dynamics.} Sketch of the fundamental processes of the quantum RD dynamics. We consider a quantum bosonic chain where each site (circles) can be occupied by an integer number $\mathrm{n} \in \mathbb{N}$ of bosons. The grayscale filling of circles pictorially represents the higher (darker) or lower (brighter) occupancy of a lattice site. White circles represent empty lattice sites. In the top line, coherent hopping according to $H$ in Eq.~\eqref{eq:qrd_hamiltonian} at rate $\Omega$ is sketched. The latter replaces classical-incoherent hopping (diffusion in the continuum limit). The reactions are irreversible and are modelled by the jump operators Eqs.~\eqref{eq:annihilation} - \eqref{eq:branching} of the Lindblad dynamics. Coagulation, branching and decay are sketched. Note that coagulation requires at least two particles on the lattice site (transition from darker-grey to brighter-grey circle), while decay can act also on single occupied site (transition from grey to white circle). In the middle line, the annihilation decay \eqref{eq:annihilation} at distance $d=1$, for the sake of illustrative purposes, is shown. For $\theta \neq 0,\pi/2$ coherent superposition between different quantum mechanical states is introduced into the dynamics. In the bottom line, the action of the jump operator \eqref{eq:second_annihilation} is sketched ($d=1$ again). In this case, superposition of three different states is introduced.} 
    \label{fig:setup}
\end{figure}
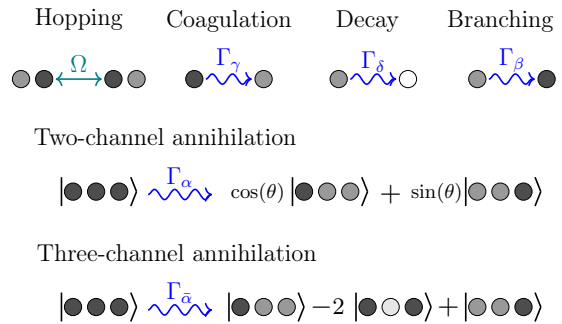

We consider bosonic particles on a quantum chain of length $L$ with periodic boundary conditions. The $j$-th lattice site of the chain can therefore either be occupied, $\hat{n}_{j} \ket{\cdots \mathrm{n}_{j} \cdots} =\mathrm{n}_{j} \ket{\cdots \mathrm{n}_{j} \cdots}$, or empty, $\hat{n}_{j} \ket{\cdots 0_{j} \cdots} = 0$, with $\hat{n}_{j} = \hat{b}_{j}^{\dagger} \hat{b}_{j}$ the number operator at site $j$. Here, $\hat{b}_{j}$ and $\hat{b}_{j}^{\dagger}$ are bosonic destruction and creation operators, respectively, obeying the bosonic commutation rule $[\hat{b}_{i},\hat{b}_{j}^{\dagger} ] = \delta_{i,j}$. Each site can be occupied by $\mathrm{n}_{j}$ bosons with $\mathrm{n} \in \mathbb{N}$, so that the total number of particles at each lattice site is unrestricted. The quantum reaction-diffusion dynamics of the system is ruled by the Lindblad master equation \cite{lindblad1976, gorini1976, breuer2002} (we set $\hbar=1$ in the whole manuscript)
\begin{equation} \label{eq:lindblad_master}
    \Dot{\rho}(t) = -i \left[H, \rho(t) \right] + \mathcal{D}[\rho (t)],
\end{equation}
where $\rho$ is the density matrix. In the quantum formulation of the RD dynamics, classical-incoherent hopping (diffusion in the continuum limit) is replaced by coherent hopping of particles between neighbouring sites according to the Hamiltonian $H$:
\begin{equation} \label{eq:qrd_hamiltonian}
    H = -\Omega \sum_{j=1}^{L} \left( \hat{b}_{j}^{\dagger}\hat{b}_{j+1} + \hat{b}_{j+1}^{\dagger}\hat{b}_{j} \right),
\end{equation}
with the hopping rate $\Omega$. The Hamiltonian, thus, conserves the number of particles ($[H, N(t)]= 0$, with $\hat{N}=\sum_{j} \hat{n}_{j}$ the total particle number). We note that $H$ describes the noninteracting Bose gas (vanishing interaction strength) on a optical lattice. The dissipator $\mathcal{D}[\rho]$ accounts for the irreversible reaction processes and it is of Lindblad form
\begin{equation} \label{eq:qrd_dissipator}
    \mathcal{D}(\rho) =  \sum_{j,\nu} \left( L_{j}^{\nu} \rho L_{j}^{\nu \dagger} - \frac{1}{2} \left\{ L_{j}^{\nu \dagger} L_{j}^{\nu},\rho \right\} \right),
\end{equation}
where $L_{j}^{\nu}$ are dubbed jump operators. The superscript $\nu$ labels the different reaction types. We consider various reaction types, which are pictorially represented in Fig.~\ref{fig:setup}. All these dissipative processes violate particle number conservation. Distance-selective binary annihilation $A+A\to \emptyset$ (rate $\Gamma_{\alpha}$) destroys pairs of particles at a distance $d$ and is represented by the jump operator
\begin{equation} \label{eq:annihilation}
    L_{j}^{\alpha} = \sqrt{\Gamma_{\alpha}} \hat{b}_{j} \left( \cos(\theta) \hat{b}_{j+d} - \sin(\theta)\hat{b}_{j-d}\right),
\end{equation}
Here, the parameter $\theta \in [0,\pi)$ allows for interferences between two decay channels. At $\theta=0$ or $\theta=\pi/2$, the annihilation operator reduces to the classical-incoherent binary annihilation of particles at distance $d$. Such incoherent distance-selective losses have been analyzed in Refs.~\cite{lossth8,lossth8bis} for hard-core bosons, where they were implemented via laser-exciting highly excited electronic Rydberg states in the facilitation regime. For $\theta\neq 0,\pi/2$, on the other hand, the jump operators \eqref{eq:annihilation} effectuate transitions into quantum superposition states, thereby creating quantum coherence. Such setting, whose microscopic origin is the quantum interference of two annihilation channels, was previously studied in Refs.~\cite{lossth1,QRD20222} for fermions. 

To further study the impact of interferences onto the quantum bosonic RD dynamics, we consider a binary annihilation process (rate $\Gamma_{\Bar{\alpha}}$) which couple three decay channels
\begin{equation} \label{eq:second_annihilation}
    L_{j}^{\Bar{\alpha}} = \sqrt{\Gamma_{\bar{\alpha}}} \hat{b}_{j} \left( \hat{b}_{j+d} + \hat{b}_{j-d} - 2 \hat{b}_{j}\right), 
\end{equation}
between particles at fixed distance $d$. We refer to the jump operator \eqref{eq:second_annihilation} henceforth as second-order annihilation reaction as $L_j^{\Bar{\alpha}}$ is constructed such that in the continuum space limit, the zeroth and the first order of the expansion the jump operator in the lattice spacing vanish (see Appendix \ref{app:cont_limit} for the details). We anticipate (see the discussion in Subsecs.~\ref{subsec:annihilation_inteferences} for further details) that both the jump operator \eqref{eq:annihilation}, with $\theta=\pi/4$, and \eqref{eq:second_annihilation} have a zero-momentum Bose-Einstein condensate state (BEC) as an exact dark state. Similar jump operators having the BEC as a many-body dark state, albeit still conserving the total particle number, have been studied, e.g., in Refs.~\cite{diehl2008,tomadin2011}.

Other types of reactions we are considering is onsite coagulation $A+A \to A$ (rate $\Gamma_\gamma$)
\begin{equation}
\label{eq:coagulation}
    L_{j}^{\gamma} = \sqrt{\Gamma_{\gamma}} \hat{n}_j\hat{b}_{j}.
\end{equation}
and one-body decay $A\to \emptyset$ (rate $\Gamma_{\delta}$)
\begin{equation}
L_j^{\delta}=\sqrt{\Gamma_{\delta}}\hat{b}_j.
\label{eq:1b_decay}
\end{equation}
In all the cases \eqref{eq:annihilation}-\eqref{eq:1b_decay}, the density of particles $n(t) = \braket{\hat{N}(t)}/L$ is a monotonically decreasing function of time $t$. The universal behavior of the dynamics lies in the asymptotic late-time approach of the system towards the steady state devoid of particles.

In order to have a nontrivial steady state, supporting a nonzero density of particles, we need to include reactions that increase the particle number (see Fig.~\ref{fig:setup}). We do this by introducing onsite branching reactions $A\to A+A$ (rate $\Gamma_\beta$)
\begin{equation} \label{eq:branching}
    L_{j}^{\beta} = \sqrt{\Gamma_{\beta}} \hat{b}_{j}^{\dagger} \hat{n}_{j},
\end{equation}
which lead to the creation of an offspring from a lattice site already occupied by at least one particle. The form of the jump operators \eqref{eq:coagulation} and \eqref{eq:branching} is dictated by bosonic statistics, allowing reactions to take place at the same lattice site. 

The competition between branching \eqref{eq:branching} and decay \eqref{eq:1b_decay} in systems with restricted occupation of lattice sites (spin or fermionic systems) leads to a second-order (continuous) stationary-state phase transitions separating an active (nonzero density) phase from an inactive-aborbing one (with zero density) \cite{Paessens_2004}. 
For bosons the maximum occupation number per site is unrestricted. Therefore, in order to have a finite density in the active phase also coagulation \eqref{eq:coagulation} (or annihilation \eqref{eq:annihilation}) has to be considered \cite{tauber2005applications,Malte_Henkel_2004,Paessens_2004}. Here, universal behavior emerges near the critical point of the ensuing second-order phase transition, as we recall in the next Sec.~\ref{sec:rd_dynamics}.

\section{Diffusion and Reaction-limited Dynamics} \label{sec:rd_dynamics}
The dynamics of the density $n(t)$ as a function time is the central quantity characterizing the emergent universal behavior of RD systems. Two different timescales rule the RD dynamics: the reaction time $\sim \Gamma^{-1}$ and the diffusion time $\sim \Omega^{-1}$. The former quantifies the time (on average) two nearby particles take to react, while the latter the time (on average) two particles take to get in contact. The dynamics changes qualitatively depending on the ratio $\Gamma/\Omega$. The limit $\Gamma/\Omega \ll 1$ is named ``reaction-limited regime'', while the limit where $\Gamma/\Omega$ is at least $\Gamma/\Omega \sim 1$, is dubbed ``diffusion-limited regime''. The exponent $\delta$ of the asymptotic power-law decay  $n(t) \sim t^{-\delta}$ is different in the two regimes. In Subsec.~\ref{subsec:classical_rd_dynamics}, we first briefly recall and discuss previous results concerning the bosonic RD dynamics for classical systems. In Subsec.~\ref{subsec:quantum_rd_dynamics}, we then recall aspects of the quantum RD dynamics that are relevant for the treatment of the present manuscript. In particular, we focus on the quantum-reaction limited regime, which is the main focus of this manuscript.

\subsection{Classical RD Dynamics} \label{subsec:classical_rd_dynamics}
Within the classical RD realm, the reaction-limited regime is well-described by the so-called ``law of mass action'' rate equation \cite{hinrichsen2000non,vladimir1997nonequilibrium,Redner1984, fastdiffusion1992,krapivsky2010kinetic}. This equation states that the rate of reactions equals the product of the concentrations of reactants. For the annihilation reaction, where $m$ particles react to $l$ particles (with $m >l \in \mathbb{N}$), the law of mass action equation yields,  
\begin{equation} 
\label{eq:rd_rl}
    \frac{\mbox{d}}{\mbox{d}t} n(t) = - (m-l) \Gamma n^{m}(t).
\end{equation}
Equation \eqref{eq:rd_rl} is valid whenever reactions occur everywhere in space with the same probability, i.e., it is valid in a well-stirred regime where the density $n(t)$ is homogeneous. This is precisely the reaction-limited regime $\Gamma/\Omega \ll 1$, where the fast hopping (diffusion in continuum space) mixing wipes out spatial patterns in the density profile coming from local depletion of the particle number. For $m \ge 2$, the long-time behaviour associated to \eqref{eq:rd_rl} is 
\begin{equation} \label{eq:rl_crd}
    n(t) \quad \propto \quad  \left( \Gamma t \right)^{-1/(m-1)}.
\end{equation}
The exponent of the power-law decay depends on the number of reactants $m$ involved in the reaction. Note that in this regime, the density is a function of the rescaled time $\tau=\Gamma t$ according to the reaction rate $\Gamma$. This further shows that in the reaction-limited regime the limiting factor for the asymptotic decay of the density is the reaction time (controlled by $\Gamma^{-1}$). 

The diffusion-limited regime, $\Gamma/\Omega \sim 1$, is qualitatively different. In this regime the hopping mixing is finite and therefore spatial fluctuations in the density profile, due to local depletion of the particle number, are relevant in the kinetics. For classical systems with unrestricted occupation, this problem has been analytically tackled in Refs.~\cite{Racz1985,tauber2002dynamic,tauber2005applications,tauber2014critical,doi1976stochastic,peliti1985path,peliti1986renormalisation,QFT_RD_1998} by mapping the classical microscopic master equation into a bosonic second-quantization problem via the Doi-Peliti formalism \cite{doi1976,doi1976stochastic}. From the latter, a field-theory description can be eventually derived. The associated renormalization group scaling analysis yields for $2A\to\emptyset$ in one dimension $D=1$: 
\begin{equation} 
\label{eq:dl_crd}
    n(t) \propto 
    (\Omega t)^{-1/2}.
\end{equation}
Here, the density depends on the rescaled time $\tau=\Omega t$ according to the diffusion rate. The decay \eqref{eq:dl_crd} applies classically also for coagulation $2A\to A$, which belongs, indeed, to the same universality class of annihilation \cite{henkel1995equivalences,henkel1997reaction,krebs1995finite,simon1995concentration,ben2005relation}. The result \eqref{eq:dl_crd} shows that the limit factor for the decay of the density of particles is, in this case, the time needed for two far apart particles to meet. This is a universal property of the random walk, which is recurrent only for $D\leq 2$. This explain both the universal character of Eq.~\eqref{eq:dl_crd} and the emergence of the upper critical dimension $D_c=2$. For $D\geq D_c$, diffusion is effective in filling the whole space and one recover the mean-field prediction \eqref{eq:rl_crd} (up to a logarithmic correction at $D=D_c=2$).

Reaction-diffusion model can also host stationary state phase transitions between an active phase and an absorbing one. This happens when branching reactions $A \to A+A$ are included. The CP model, namely, describes the stationary phase transition ensuing from the competition between branching \eqref{eq:branching} and \eqref{eq:1b_decay}. In the case where site occupancy is unrestricted, the so-called bosonic contact process \cite{Malte_Henkel_2004,Paessens_2004}, one further needs to include coagulation \eqref{eq:coagulation} (or annihilation \eqref{eq:annihilation} at $\theta=0$ and $d=0$) in order to have a finite stationary density. This can be heuristically understood by writing the law of mass action\eqref{eq:rd_rl} in the presence of branching and decay
\begin{equation}
\frac{\mbox{d}n(t)}{\mbox{d}t}=(\Gamma_{\beta}-\Gamma_{\delta})n(t)-\Gamma_{\gamma}n^2(t).
\label{eq:MF_contact_process}
\end{equation}
This equation trivially admits the stationary solution $n_{\mathrm{stat}}=0$ corresponding to the absorbing phase. Furthermore, one has the additional solution
\begin{equation}
n_{\mathrm{stat}}=\frac{\Gamma_{\beta}-\Gamma_{\delta}}{\Gamma_{\gamma}},
\label{eq:MF_CP_bosons}
\end{equation}
if $\Gamma_{\beta}>\Gamma_{\delta}$. In the case of restricted CP, one does not need coagulation to have a finite stationary density since a term proportional to $n^2$ is produced on the right hand side of \eqref{eq:MF_contact_process} due to the restricted occupation constraint [$n(t)(1-n(t))$ from the branching]. Equation \eqref{eq:MF_contact_process} immediately locates the mean-field critical point $\Gamma^c_{\beta}=\Gamma_{\delta}$ of the phase transition. This transition is classically of second order and it belongs to the directed percolation universality class \cite{ vladimir1997nonequilibrium, henkel2008non, hinrichsen2000non,krapivsky2010kinetic,tauber2014critical,Malte_Henkel_2004}. Universal behavior lies in the algebraic behavior of the order parameter $n_{\mathrm{stat}} \sim (\Gamma_{\beta}-\Gamma^{c}_{\beta})^{\beta}$ close to the transition point. The associated critical exponent is $\beta=1$ within the reaction-limited regime. Away from this limit, the exponent $\beta \neq 1$ differs from the mean-field prediction and it depends on the space-dimensionality for $D\leq D_c$, and $D_c=4$ \cite{vladimir1997nonequilibrium, henkel2008non, hinrichsen2000non,krapivsky2010kinetic,tauber2014critical,Malte_Henkel_2004}. We do not report these values here as they are not central for the understanding of the results of this manuscript, which is focused on the quantum analogue of Eqs.~\eqref{eq:MF_contact_process} and \eqref{eq:MF_CP_bosons}. We introduce the quantum reaction-limited regime in the next Subsection. We will see in Sec.~\ref{sec:results} that the ensuing quantum reaction-limited equation for the CP displays a much richer phase diagram than that of Eqs.~\eqref{eq:MF_contact_process} and \eqref{eq:MF_CP_bosons}.

\subsection{Quantum RD Dynamics} \label{subsec:quantum_rd_dynamics}
For quantum RD systems little is currently known. The quantum master equation in Eqs.~\eqref{eq:lindblad_master}-\eqref{eq:branching} cannot be solved analytically since it is not quadratic as a consequence of the structure of the jump operators. At the same time, large-scale numerics, involving large system sizes and long times, is severely hindered since the simulation of the quantum trajectories associated to Eq.~\eqref{eq:lindblad_master} requires the knowledge of the whole many-body wavefunction. Due to the exponential scaling of the Hilbert space with the system size $L$ such simulations become rapidly infeasible. For example, in Ref.~\cite{RDHorssen}, the quantum XX spin-chain Hamiltonian (which maps to free fermions via Jordan-Wigner transformation) subject to binary annihilation $2A\to\emptyset$ or coagulation $2A\to A$ has been studied. Therein the diffusion-limited regime $\Omega=\Gamma=1$ is considered, the initial state is the fully filled product state $\ket{\bullet \bullet \dots \bullet}$ and a maximal system size of $L=22$ is taken. Under such conditions, a power-law decay $n(t)\sim t^{-b}$, with $1/2<b<1$ is obtained for annihilation. The dynamics is therefore slower than the mean-field prediction \eqref{eq:rd_rl} but faster than the classical diffusion-limited analogue \eqref{eq:dl_crd}. This might be related to the faster ballistic spreading of quantum particles, compared to classical diffusion, albeit the extrapolation of these results to the thermodynamic limit is difficult to assess. No analytical prediction for the quantum-diffusion limited decay exponent is far present. A field-theory description for $2A\to\emptyset$ has been proposed in Ref.~\cite{gerbino2023large} using the Keldysh field-theory representation of the quantum master equation. A systematic renormalization group scaling analysis of this action is, however, still missing.

The quantum reaction-limited regime ($\Gamma/\Omega \ll 1$) is in this perspective unique since it allows exact analytical calculations in the thermodynamic limit and at long times. This regime has been recently studied in a number of works for fermionic and spin systems \cite{lossth1,lossth2,lossth6,QRD20222,perfetto2023quantum,riggio2023effects}, where occupation restrictions are therefore present. It has been therein shown that for fermionic particles the binary annihilation ($2A \to \emptyset$) decays as $n(t) \sim t^{-1/2}$ for quantum coherent (in real space) initial states. The dynamics is therefore not always well-described by the mean-field approach despite the system being in the reaction-limited regime. Quantum effects due to coherences therefore allow for novel form of emergent behavior, which do not have a classical counterpart. The interacting Bose gas, modelled by the Lieb-Liniger Hamiltonian, subject to $m$-body annihilation, $mA\to\emptyset$, has been considered in Ref.~\cite{lossth3}. In the noninteracting limit, mean-field decay \eqref{eq:rd_rl} is found for every $m$. In the presence of nonzero interaction, the numerical evaluation of the differential equation for the density turns out to be cumbersome and therefore no estimate for the density decay exponent is therein made. The analysis of all the works \cite{lossth1,lossth2,lossth3,lossth6,QRD20222,perfetto2023quantum,riggio2023effects} is based on the analytical approach of the time-dependent Gibbs ensemble (TGGE) method \cite{tGGE1,tGGE2,tGGE3,tGGE4}. We briefly recall here this method as it will be used for all the results presented in Sec.~\ref{sec:results}.

The reaction-limited regime $\Gamma/\Omega \ll1$ corresponds to weak dissipation regime of the Lindblad dynamics. In this limit, the reaction time $\sim 1/\Gamma$ is much larger than the hopping time $\sim 1/\Omega$. 
Because of this separation of time scales, the state of the system $\rho(t)$ quickly relaxes to a stationary state $\rho_{SS}(t)$ of the Hamiltonian $\left[H , \rho_{SS}(t) \right]=0$. The stable quasi-particles characterizing the integrable Hamiltonian $H$ \eqref{eq:qrd_hamiltonian} (it is trivially integrable since it is quadratic) can still be defined but they acquire a finite lifetime $\sim 1/\Gamma$. This causes the stationary state of the system becoming time-dependent $\rho_{SS}(t)$ and changing over the long time scale $1/\Gamma$. The TGGE method makes an ansatz for $\rho_{SS}(t)$ in the form of a GGE  \cite{GGErev1,GGErev2}.This ansatz is motivated by the idea of local generalized thermalization of the Hamiltonian dynamics in the limit of slow reactions. In this limit, the density matrix quickly relaxes towards a maximal-entropy generalized Gibbs state, which keeps into account all the conserved charges of the Hamiltonian. In the case of the Hamiltonian \eqref{eq:qrd_hamiltonian}, all the conserved charges are linear combination of the occupation number $\hat{n}_k$ in momentum space \cite{GGErev1,GGErev2}. For the Hamiltonian \eqref{eq:qrd_hamiltonian}, the time-dependent GGE (TGGE) then reads as \cite{tGGE1, tGGE2,tGGE3,tGGE4} 
\begin{equation} 
\label{eq:tGGE}
    \rho_{\mathrm{GGE}} (\tau) = \frac{1}{\mathcal{Z}(\tau)} \exp \left(- \sum_{k} \lambda_{k}(\tau) \hat{n}_{k}\right),
\end{equation}
with $\mathcal{Z}(\tau) = \mathrm{Tr}[\exp(- \sum_{k} \lambda_{k}(\tau) \hat{n}_{k})]$. Here $k\in (-\pi,\pi)$ is the quasi-momentum and $\hat{n}_{k} = \hat{b}_{k}^{\dagger} \hat{b}_{k}$ the number operator in Fourier space with $\hat{b}_{k},\hat{b}_k^{\dagger}$ the bosonic Fourier-transformed operators (see Appendix \ref{app:binary_ann}). The GGE state describes averages $\braket{\mathcal{O}}_{\mathrm{GGE}}$ of local observables $\mathcal{O}$ in the thermodynamic limit. The time-dependence of the TGGE is contained in the Lagrange multipliers (or generalised inverse temperatures) $\lambda_{k} \to \lambda_{k}(\tau)$. This implies that conserved charges can still be defined, but they slowly (on a time scale $\sim 1/\Gamma$) drift in time as consequence of the dissipation. Since $[\hat{n}_{k}, H]=0$ for all $k$, $[\rho_{\mathrm{GGE}}(t),H]=0$ and the master equation \eqref{eq:lindblad_master} takes the form:
\begin{equation} \label{eq:master_tgge}
    \frac{d}{dt} \rho_{\mathrm{GGE}} (t) = \mathcal{D}[\rho_{\mathrm{GGE}}(t)].
\end{equation}
The state in Eq.~\eqref{eq:tGGE} is diagonal in momentum space (because of translation invariance) and Gaussian. The entire dynamics ensuing from \eqref{eq:tGGE} is consequently encoded in the bosonic occupation function in momentum space $B_{q} (t)\equiv \braket{\hat{b}_q^{\dagger}\hat{b}_k}_{\mathrm{GGE}}(t) = \delta_{q,k}/(\exp(\lambda_{q}(t))-1)$. The time evolution for the Lagrange parameters $\lambda_q(t)$ is therefore in one-to-one correspondence with that of $B_q(t)$. The dynamical differential equation for the latter immediately follows from Eq.~\eqref{eq:master_tgge} exploiting that $[\rho_\mathrm{GGE},\hat{n}_\mathrm{q}]=0$ \cite{QRD20222, lossth1, lossth2, lossth6}:
\begin{equation}
\label{eq:qrd_diff}
    \frac{d B_{q}(t)}{dt} = \sum_{j, \nu} \braket{L_{j}^{\nu \dagger} \left[ \hat{n}_{q}, L_{j}^{\nu} \right]}_{\mathrm{GGE}},  \quad \forall q.
\end{equation}
In the Appendices \ref{app:binary_ann}-\ref{app:branching_bosons}, we report the expression of the jump operators $L_j^{\nu}$ in Fourier space and we detail all the calculations based on Eq.~\eqref{eq:qrd_diff}.

It is also clear from the previous equation that $B_q(t)=B_q(\tau=\Gamma t)$, i.e., the bosonic occupation function is a function of the rescaled time $\tau=\Gamma t$ according to the reaction rate only. This is consistent with the reaction-limited description of the classical RD dynamics \eqref{eq:rd_rl} and \eqref{eq:rl_crd} where time obeys the very same rescaling. As a matter of fact, the TGGE ansatz of Eqs.~\eqref{eq:tGGE} and \eqref{eq:qrd_diff} is valid in the scaling limit $t\to\infty$, $\Gamma \to 0$ with $\tau=\Gamma t$ fixed, as shown in Refs.~\cite{tGGE1, tGGE2,tGGE3,tGGE4} We also mention that in the present case $B_q(\tau)$ does not depend on the space coordinate $x$ since we only deal in this manuscript with homogeneous Lindbladians evolving from homogeneous initial states. This is also the reason why the hopping rate $\Omega$ does not appear in the equation \eqref{eq:qrd_diff}. In homogeneous systems, indeed, there is no transport of particles. Consequently, in the limit $\Gamma/\Omega \ll 1$, the Hamiltonian contribution is effectively integrated out assuming local relaxation to the GGE state \eqref{eq:tGGE}. Whenever spatial inhomogeneities are included, the Hamiltonian additionally contributes to Eq.~\eqref{eq:master_tgge} via a convective term describing ballistic transport of particles. Equation \eqref{eq:master_tgge} takes in this case the form of a Boltzmann equation \cite{lossth2,lossth3,riggio2023effects,gerbino2023large}.

The right hand side of Eq.~\eqref{eq:qrd_diff} can be exactly computed from the Wick's theorem since it amounts to computing higher-point bosonic correlation functions over the Gaussian state \eqref{eq:tGGE}. This allows to derive exact rate equations for $B_q(\tau)$. Within this perspective, the TGGE method presented here bears similarities with the Hartree-Fock decoupling of four (and higher) body terms in the dynamical equation for the two-point function $\braket{\hat{b}^{\dagger}_n \hat{b}_m}$ (with $m,n$ lattice-site indices). This decoupling is often used in dissipative many-body systems, see, e.g., Ref.~\cite{lossthHFdecoupling}, in order to truncate the infinite hierarchy of equations for the correlation functions to a closed and calculable form. We, however, remark that the Hartree-Fock decoupling is an uncontrolled approximation, while the validity of the TGGE ansatz is controlled by the parameter $\Gamma \to 0$ (reaction-limited regime) according to the scaling limit $\tau=\Gamma t$ fixed explained above. Furthermore, the similarity between the two methods applies only in the present case of the noninteracting Bose gas with quadratic Hamiltonian \eqref{eq:qrd_hamiltonian}. As discussed in Ref.~\cite{lossth3}, for the interacting Bose gas described by the Lieb-Liniger Hamiltonian, the GGE is not Gaussian and higher point function in the GGE state cannot be reduced to the two-point function via Wick theorem. In the next Section, we specialize \eqref{eq:qrd_diff} for the various reaction processes introduced before in Eqs.~\eqref{eq:annihilation}-\eqref{eq:branching}.   

\section{Results} \label{sec:results}
In this Section, we present our results for the quantum reaction-limited RD dynamics of bosonic particles. In Subsec.~\ref{subsec:distance_selective_loss}, we first discuss the distance-selective annihilation process \eqref{eq:annihilation} ($\theta=0$). In Subsec.~\ref{subsec:annihilation_inteferences}, we move to the discussion of annihilation processes with interferences between two decay channels, \eqref{eq:annihilation} with $\theta \neq 0,\pi/2$. In Subsec.~\ref{subsec:second_order_annihilation}, we discuss the second-order annihilation process with three interfering decay channels \eqref{eq:second_annihilation}. In Subsec.~\ref{subsec:coagulation}, coagulation decay \eqref{eq:coagulation} is studied. In Subsec.~\ref{subsec:branching}, we consider the branching process \eqref{eq:branching}. We then study the corresponding absorbing-state phase transition arising from the competition between branching \eqref{eq:branching}, coagulation \eqref{eq:coagulation} and one-body decay \eqref{eq:1b_decay}. The associated stationary phase diagram is reported and discussed.

For all the aforementioned cases, we solve the TGGE rate equation in Eq.~\eqref{eq:qrd_diff} for the bosonic momentum occupation function $B_q(\tau)$. Namely, we consider three different initial conditions: the Bose Einstein condensate (BEC) , Eq.~\eqref{eq:bc}, the flat mode filling, Eq.~\eqref{eq:ff}, and the Gaussian distributed mode filling, Eq.~\eqref{eq:gd}. The implemented initial conditions are the following:
\begin{itemize}
    \item Bose Condensate:
    \begin{equation} \label{eq:bc}
        B_q(0)=B_{\mathrm{BEC},q}= N \cdot \delta_{q, 0},
    \end{equation}
    \item Flat Filling:
    \begin{equation} \label{eq:ff}
        B_q(0)=B_{\mathrm{FF}} = n_{0},
    \end{equation}
    \item Gaussian State:
    \begin{equation} \label{eq:gd}
       B_q(0)=B_{\mathrm{GS},q} =  2\sqrt{\pi} \cdot e^{-q^{2}}.
    \end{equation}
\end{itemize}
In the BEC, the quasi-momentum $q=0$ is macroscopically populated by the total number $N$ of bosons initially in the system. In one spatial dimension, this state corresponds to a quasi-condensate pure state $\ket{\mathrm{BEC}}=(\hat{b}^{\dagger}_{q=0})^{N}\ket{0}/\sqrt{N!}$ \cite{pethick2008bose,cazalilla2011}. In the flat filling, every mode in momentum space is equally occupied by the same number $n_{0}$ of bosons. The parameter $n_{0}$ is therefore the initial density of particles. In the state \eqref{eq:gd}, the occupation of the modes is Gaussian distributed. The density is computed from the occupation function $B_{q}(\tau)$ as
\begin{equation}
    \braket{n(\tau)}_{\mathrm{GGE}} = \frac{1}{L} \sum_{q} B_{q}(\tau)=\int_{-\pi}^{\pi}\frac{\mbox{d}q}{2\pi}B_q(\tau),
\label{eq:density_integral}    
\end{equation}
the latter equality being valid as $L\to\infty$. Henceforth, we set $N=L$ in Eq.~\eqref{eq:bc}, and $n_{0}=1$ in Eq.~\eqref{eq:ff}, so that for all the three initial conditions we have an initial density $n_0=1$ \footnote{For the state \eqref{eq:gd}, the integral \eqref{eq:density_integral} is only approximately equal to 1 since the integral is restricted on the lattice to the Brillouin zone $k \in (-\pi,\pi)$.}.

The BEC state \eqref{eq:bc} has quantum coherences in real space since the associated density matrix has non-zero off-diagonal matrix elements in the bosonic Fock-space basis. The latter being spanned by states of the form $\ket{\{\mathrm{n}_j\}}=\prod_{j \in \Lambda} \hat{b}_j^{\mathrm{n}_j}\ket{0}$, with $\Lambda$ denoting an arbitrary set of lattice sites. Similarly, the Gaussian state \eqref{eq:gd} corresponds to an initial state of the GGE form \eqref{eq:tGGE} identified by the occupation function $B_{\mathrm{GS},q}$ inhomogeneous in momentum space. The latter is associated to a bosonic two-point correlation function in real space $\braket{\hat{b}^{\dagger}_n \hat{b}_m}_{\mathrm{GGE}}$ not diagonal. The latter fully characterizes the Gaussian initial state, which is therefore also in this case quantum coherent in the real space basis. On the contrary, for the flat initial occupation in momentum space \eqref{eq:ff}, the two-point bosonic correlation function is diagonal. The associated initial density matrix $\rho_0= \mbox{exp}(-\lambda N)/\mathcal{Z}$ is diagonal in the classical basis of the Fock space introduced above. For these reasons, we will refer henceforth to initial states \eqref{eq:bc} and \eqref{eq:gd} as quantum coherent, in contrast with \eqref{eq:ff} which is incoherent.

The asymptotic decay exponent $\braket{n}_{\mathrm{GGE}}(\tau) \sim \tau^{-\delta}$ is evaluated by computing the effective exponent \cite{hinrichsen2000non}:
\begin{equation} \label{eq:eff_exponent}
     \delta_{\text{eff}}(\tau) = -\frac{\log \left(\braket{n}_{\mathrm{\mathrm{GGE}}}(b \tau)/\braket{n}_{\mathrm{GGE}}(\tau)\right)}{\log(b)}. 
\end{equation}
Here, $b>0$ is a scaling parameter. In all the calculations, we set $b=1.5$. In the case of a power-law decay, the effective decay exponent $\delta_{\text{eff}}(\tau)$ approaches asymptotically in time the exponent $\delta$. We plot in the next Subsections $\delta_{\text{eff}}$ as a function of $\tau$ for the three different initial states \eqref{eq:bc}-\eqref{eq:gd} for the various reactions considered \eqref{eq:annihilation}-\eqref{eq:coagulation}. 

\subsection{Distance-selective loss} \label{subsec:distance_selective_loss}
\begin{figure}[t]
\centering     
    \centering
    \raggedright\text{\small a)}
    \vfill
    \resizebox{0.465\columnwidth}{!}{\tikzsetnextfilename{distances_flat}
\begin{tikzpicture}
\begin{axis}[
            xmode=log,
            ymode=log,
            ylabel={$\braket{n}_{GGE}(\tau)$},
            xlabel={$\tau$},
            legend style={at={(0.6,0.3)}, anchor=east},
            every tick/.style={black,semithick},
            ]
        \addplot[color=red,smooth,thick, scatter src=explicit symbolic] table[col sep=comma,header=true,x index=1,y index=2] {Results/Plots_A/Data/ann_distance_flat.csv}; \addlegendentry{$d = 0$}
        \addplot[color=blue,smooth,very thick, scatter src=explicit symbolic] table[col sep=comma,header=true,x index=1,y index=3] {Results/Plots_A/Data/ann_distance_flat.csv}; \addlegendentry{$d = 1$}
        \addplot[color=orange, smooth,very thick,  scatter src=explicit symbolic] table[col sep=comma,header=true,x index=1,y index=4] {Results/Plots_A/Data/ann_distance_flat.csv}; \addlegendentry{$d = 20$}
\end{axis}
\end{tikzpicture}}
    \hfill
    \resizebox{0.435\columnwidth}{!}{\tikzsetnextfilename{x_distances_flat_exponent}
\begin{tikzpicture}
\begin{axis}[
            xmode=log,
            ylabel={ $\delta_{\mathrm{eff}}(\tau)$},
            xlabel={$\tau$},
            ytick={0,0.5,1},
            every tick/.style={black,semithick},
            minor y tick num=1,
            xmax=1e5
            ]
        \addplot[color=red,smooth,thick, scatter src=explicit symbolic] table[col sep=comma,header=true,x index=1,y index=2] {Results/Plots_A/Data/distances_flat_exponent.csv};
        \addplot[color=blue,smooth,very thick, scatter src=explicit symbolic] table[col sep=comma,header=true,x index=1,y index=3] {Results/Plots_A/Data/distances_flat_exponent.csv};
        \addplot[color=orange, smooth,very thick,  scatter src=explicit symbolic] table[col sep=comma,header=true,x index=1,y index=4] {Results/Plots_A/Data/distances_flat_exponent.csv};
\end{axis}
\end{tikzpicture}}
    \vfill
    \raggedright{\text{\small b)}}
    \vfill
    \resizebox{0.465\columnwidth}{!}{\tikzsetnextfilename{distances_gauss}
\begin{tikzpicture}
\begin{axis}[
            xmode=log,
            ymode=log,
            ylabel={$\braket{n}_{GGE}(\tau)$},
            xlabel={$\tau$},
            ytick={0.000001, 0.0001, 0.01, 1},
            xtick={10^-3,10^0,10^3,10^6},
            every tick/.style={black,semithick},
            ]
        \addplot[color=red,smooth,thick, scatter src=explicit symbolic] table[col sep=comma,header=true,x index=1,y index=2] {Results/Plots_A/Data/ann_distance_gauss_2.csv};
        \addplot[color=blue,smooth,very thick, scatter src=explicit symbolic] table[col sep=comma,header=true,x index=1,y index=3] {Results/Plots_A/Data/ann_distance_gauss_2.csv};
        \addplot[color=orange, smooth,very thick,  scatter src=explicit symbolic] table[col sep=comma,header=true,x index=1,y index=4] {Results/Plots_A/Data/ann_distance_gauss_2.csv};
\end{axis}
\end{tikzpicture}}
    \hfill
    \resizebox{0.435\columnwidth}{!}{\tikzsetnextfilename{distances_gauss_exponent}
\begin{tikzpicture}
\begin{axis}[
            xmode=log,
            ylabel={$\delta_{\mathrm{eff}}(\tau)$},
            xlabel={$\tau$},
            ytick={0,0.5,1},
            xtick={10^-3,10^0,10^3,10^6},
            every tick/.style={black,semithick},
            minor y tick num=1,
            xmax=1e5
            ]
        \addplot[color=red,smooth,thick, scatter src=explicit symbolic] table[col sep=comma,header=true,x index=1,y index=2] {Results/Plots_A/Data/distances_gauss_exponent.csv};
        \addplot[color=blue,smooth,very thick, scatter src=explicit symbolic] table[col sep=comma,header=true,x index=1,y index=3] {Results/Plots_A/Data/distances_gauss_exponent.csv};
        \addplot[color=orange, smooth,very thick,  scatter src=explicit symbolic] table[col sep=comma,header=true,x index=1,y index=4] {Results/Plots_A/Data/distances_gauss_exponent.csv};
\end{axis}
\end{tikzpicture}}
    \caption{\textbf{Distance-selective binary annihilation quantum RD dynamics.} (Left) Log-log plot of the density $\braket{n}_{\mathrm{GGE}}(\tau)$ as a function of $\tau=\Gamma_{\alpha}t$ from the solution of Eq.~\eqref{eq:sol_distance_selective}. (Right) Plot of the associated effective exponent $\delta_{\mathrm{eff}}(\tau)$ \eqref{eq:eff_exponent} as a function of $\tau$ (log scale on the horizontal axis only). \textbf{a)} Dynamics from the flat filling initial state \eqref{eq:ff} for three different values of the distance $d=0,1,20$. For $d=0$, the density follows Eq.~\eqref{eq:MF_onsite_results}, while for $d=1,20$ Eq.~\eqref{eq:law_of_action_d_results} is recovered. In all the cases, the asymptotic exponent (right plot) follows the mean-field prediction $\braket{n}_{\mathrm{GGE}}(\tau)\sim \tau^{-1}$. \textbf{b)} Dynamics from the initial Gaussian occupation function \eqref{eq:gd} for the same values of $d$ as above. For $d=0$, the density dynamics again follows \eqref{eq:MF_onsite_results}, while for $d=1,20$, Eq.~\eqref{eq:law_of_action_d_results} is asymptotically obtained. Also in this case the asymptotic decay exponent follows from mean field, as the convergence of $\delta_{\mathrm{eff}}(\tau)\to 1$ shows (right plot).} 
    \label{fig:ann_distances}
\end{figure}
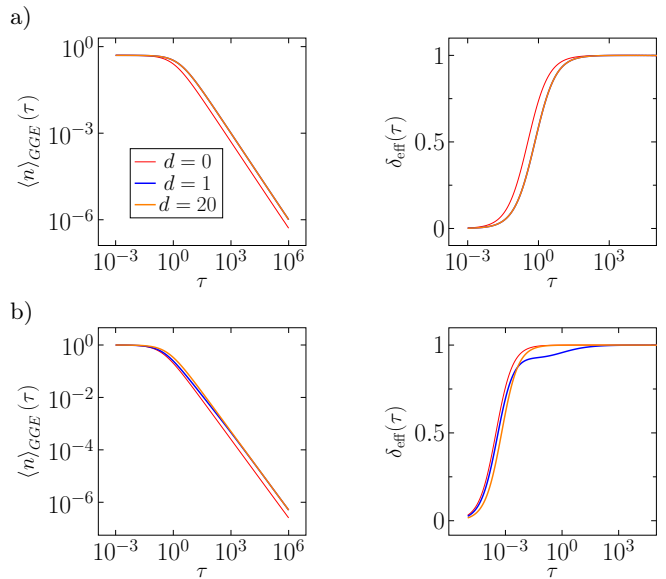

In this Subsection, we consider the binary annihilation jump operator in Eq.~\eqref{eq:annihilation} without interferences (that is $\theta = 0$). We consider generic values of the distance $d$ between the two bosons. The rate equation Eq.~\eqref{eq:qrd_diff} then yields (see Appendix \ref{app:binary_ann})
\begin{equation} \label{eq:sol_distance_selective}
    \frac{\mbox{d} B_{q}}{\mbox{d}\tau} = - \frac{2}{L} B_{q} \sum_{k} B_{k} \left( 1 + \cos(d(k-q))\right),
\end{equation}
with $\tau=\Gamma_{\alpha} t$ the re-scaled time. The results for different values of the distance $d$ are shown in Fig.~\ref{fig:ann_distances}. In particular, we see that for $d=0$ Eq.~\eqref{eq:sol_distance_selective} does not couple $q$ with other momenta $k\neq q$ and it therefore gives a closed equation for the density $\braket{n}_{\mathrm{GGE}}(t)$:
\begin{equation}
    \frac{\mbox{d} \braket{n}_{\mathrm{GGE}}(\tau)}{\mbox{d}\tau} = - 4\braket{n}_{\mathrm{GGE}}^{2}(\tau).
\label{eq:MF_onsite_results}
\end{equation}
This equation coincides with the law of mass action \eqref{eq:rd_rl} up to a factor $4$ (instead of $2$). This factor only affects the amplitude of the asymptotic decay, but not the exponent $\braket{n}_{\mathrm{GGE}}(\tau) \sim \tau^{-1}$. This result is consistent with that of Ref.~\cite{lossth3} for onsite binary losses $mA\to \emptyset$ and $m=2$. For $m>2$, in Ref.~\cite{lossth3}, the asymptotic mean-field exponent $n(\tau) \sim \tau^{-1/(m-1)}$ is also recovered.  

For $d\neq 0$, Eq.~\eqref{eq:MF_onsite_results} still holds true for the BEC initial state \eqref{eq:bc}. In this case, indeed, binary losses do not populate momenta $q\neq0$, so that the momentum distribution is concentrated on $q=0$ $B_q(\tau)=N(\tau) \delta_{q,0}$ at any time $\tau$. In this case, Eq.~\eqref{eq:sol_distance_selective} reduces to \eqref{eq:MF_onsite_results} for any $d$ value. For the flat filling initial state Eq.~\eqref{eq:ff}, instead, the occupation function $B_q(\tau)=n(\tau)$ remains flat in $q$ at any time $\tau$ and therefore \eqref{eq:sol_distance_selective} exactly reduces, at any time $\tau$, for $d\neq 0$ to the law of mass action 
\begin{equation}
  \frac{\mbox{d} \braket{n}_{\mathrm{GGE}}(\tau)}{\mbox{d}\tau} =   - 2\braket{n}_{\mathrm{GGE}}^{2}(\tau).
\label{eq:law_of_action_d_results} 
\end{equation}
This result is analogous to the one valid for fermions starting from incoherent initial states identified by a flat in $q$ occupation function \cite{QRD20222,perfetto2023quantum,riggio2023effects}. Also in those cases the fermionic quantum reaction-limited RD dynamics reduces to its classical counterpart. Meanwhile, for the initial Gaussian occupation function, studied in Refs. \eqref{eq:gd}, Eq.~\eqref{eq:sol_distance_selective} reduces to Eq.~\eqref{eq:law_of_action_d_results} only asymptotically for long $\tau$ values. The time needed to asymptotically approach \eqref{eq:law_of_action_d_results} depends on the distance $d>0$, the larger the latter, the faster the approach to the law of mass action dynamical behavior. This behavior holds generically for any initial state identified by an occupation function $B_q(0)$ inhomogeneous in momentum $q$.

Therefore, all the initial states considered show the same mean-field algebraic decay:
\begin{equation}
\label{eq:MF_decay_result_annihilation}
    \braket{n}_{\mathrm{GGE}}(\tau) \sim \tau^{-1}.
\end{equation} 
This applies to both quantum coherent \eqref{eq:bc},\eqref{eq:gd} and incoherent \eqref{eq:ff} initial conditions. This is a fundamental difference from the RD dynamics of fermions. For fermionic systems, indeed, nearest neighbour binary annihilation ($d=1$ with the notation of this manuscript) and quantum coherent initial states give rise to the decay law $\braket{n}_{\mathrm{GGE}}(\tau)\sim \tau^{-1/2}$ \cite{QRD20222}, which deviates from the mean-field result. In the next Subsection, we study whether such bosonic mean-field decay is robust against the introduction of interference effects ($\theta \neq 0,\pi/2$) in the annihilation reaction channels.

\subsection{Annihilation reaction with interferences} \label{subsec:annihilation_inteferences}
\begin{figure}[h!]
\centering  
    \centering
    \raggedright\text{\small a)}
    \vfill
    \resizebox{0.482\columnwidth}{!}{\tikzsetnextfilename{int_bc_theta}
\begin{tikzpicture}
\begin{axis}[
            xmode=log,
            ymode=log,
            ylabel={$\braket{n}_{\text{GGE}} (\tau)$},
            xlabel={$\tau$},
            legend style={at={(0.6,0.3)}, anchor=east},
            major grid style={line width=.2pt,draw=gray!50},
            enlargelimits=true, 
            ]
        \addplot[color=blue,smooth,very thick, scatter src=explicit symbolic] table[col sep=comma,header=true,x index=1,y index=3] {Data/first_order_bc_theta.csv};  \addlegendentry{$\theta = \pi/3$}
        \addplot[color=red, smooth,very thick,  scatter src=explicit symbolic] table[col sep=comma,header=true,x index=1,y index=2] {Data/first_order_bc_theta.csv};  \addlegendentry{$\theta = \pi/4$}
        \addplot[color=orange, smooth,very thick,  scatter src=explicit symbolic] table[col sep=comma,header=true,x index=1,y index=4] {Data/first_order_bc_theta.csv};  \addlegendentry{$\theta = \pi/5$}
\end{axis}
\end{tikzpicture}}
    \hfill
    \resizebox{0.458\columnwidth}{!}{\tikzsetnextfilename{int_bc_theta_exponent}
\begin{tikzpicture}
\begin{axis}[
            xmode=log,
            ylabel={$\delta_{\text{eff}} (\tau)$},
            xlabel={$\tau$},
            legend style={at={(1,0.7)}, anchor=east},
            ytick={1, 0.5,0},
            enlargelimits=true,
            minor y tick num=1, 
            ]
        \addplot[color=blue,smooth,very thick, scatter src=explicit symbolic] table[col sep=comma,header=true,x index=1,y index=3] {Data/first_order_thet_bc_exponent.csv}; 
        \addplot[color=orange, smooth,very thick,  scatter src=explicit symbolic] table[col sep=comma,header=true,x index=1,y index=4] {Data/first_order_thet_bc_exponent.csv};
\end{axis}
\end{tikzpicture}}
    \vfill
    \raggedright\text{\small b)}
    \vfill
    \resizebox{0.482\columnwidth}{!}{\tikzsetnextfilename{int_flat_theta}
\begin{tikzpicture}
\begin{axis}[
            xmode=log,
            ymode=log,
            ylabel={$\braket{n}_{\text{GGE}} (\tau)$},
            xlabel={$\tau$},
            legend style={at={(1,0.7)}, anchor=east},
            enlargelimits=true,
            ]
        \addplot[color=blue,smooth,very thick, scatter src=explicit symbolic] table[col sep=comma,header=true,x index=1,y index=3] {Data/first_order_flat_theta.csv};
         \addplot[color=red, smooth,very thick,  scatter src=explicit symbolic] table[col sep=comma,header=true,x index=1,y index=2] {Data/first_order_flat_theta.csv};
        \addplot[color=orange, smooth,very thick,  scatter src=explicit symbolic] table[col sep=comma,header=true,x index=1,y index=4] {Data/first_order_flat_theta.csv}; 
\end{axis}
\end{tikzpicture}}
    \hfill
    \resizebox{0.458\columnwidth}{!}{\tikzsetnextfilename{int_flat_theta_exponent}
\begin{tikzpicture}
\begin{axis}[
            xmode=log,
            ylabel={$\delta_{\text{eff}} (\tau)$},
            xlabel={$\tau$},
            legend style={at={(1,0.2)}, anchor=east},
            enlargelimits=true,
            ytick={1, 0.5,0},
            minor y tick num=1, 
            ]
        \addplot[color=blue,smooth,very thick, scatter src=explicit symbolic] table[col sep=comma,header=true,x index=1,y index=3] {Data/first_order_thet_flat_exponent.csv};
        \addplot[color=red, smooth,very thick,  scatter src=explicit symbolic] table[col sep=comma,header=true,x index=1,y index=2] {Data/first_order_thet_flat_exponent.csv};
        \addplot[color=orange, smooth,very thick,  scatter src=explicit symbolic] table[col sep=comma,header=true,x index=1,y index=4] {Data/first_order_thet_flat_exponent.csv};
\end{axis}
\end{tikzpicture}}
    \vfill
\raggedright\text{\small c)}
    \vfill
    \resizebox{0.482\columnwidth}{!}{\tikzsetnextfilename{int_gauss_theta}
\begin{tikzpicture}
\begin{axis}[
            xmode=log,
            ymode=log,
            ylabel={$\braket{n}_{\text{GGE}} (\tau)$},
            xlabel={$\tau$},
            xtick={10^-3,10^0,10^3,10^6, 10^9},
            legend style={at={(1,0.8)}, anchor=east},
            enlargelimits=true,
            ]
        \addplot[color=blue,smooth,very thick, scatter src=explicit symbolic] table[col sep=comma,header=true,x index=1,y index=2] {Data/first_order_gauss_theta_2.csv};
        \addplot[color=red, smooth,very thick,  scatter src=explicit symbolic] table[col sep=comma,header=true,x index=1,y index=3] {Data/first_order_gauss_theta_2.csv}; 
        \addplot[color=orange, smooth,very thick,  scatter src=explicit symbolic] table[col sep=comma,header=true,x index=1,y index=4] {Data/first_order_gauss_theta_2.csv};
\end{axis}
\end{tikzpicture}}
    \hfill
    \resizebox{0.458\columnwidth}{!}{\tikzsetnextfilename{int_gauss_theta_exponent}
\begin{tikzpicture}
\begin{axis}[
            xmode=log,
            ylabel={$\delta_{\text{eff}} (\tau)$},
            xlabel={$\tau$},
            legend style={at={(0.9,0.4)}, anchor=east},
            enlargelimits=true,
            xtick={10^-3,10^0,10^3,10^6, 10^9},
            ytick={1, 0.5,0},
            minor y tick num=1, 
            ]
        \addplot[color=blue,smooth,very thick, scatter src=explicit symbolic] table[col sep=comma,header=true,x index=1,y index=2] {Data/first_order_thet_gauss_exponent_2.csv}; 
        \addplot[color=red, smooth,very thick,  scatter src=explicit symbolic] table[col sep=comma,header=true,x index=1,y index=3] {Data/first_order_thet_gauss_exponent_2.csv}; 
        \addplot[color=orange, smooth,very thick,  scatter src=explicit symbolic] table[col sep=comma,header=true,x index=1,y index=4] {Data/first_order_thet_gauss_exponent_2.csv};
\end{axis}
\end{tikzpicture}}
    \caption{\textbf{Quantum RD dynamics of binary annihilation with interference between two decay channels} (Left) Log-log plot of the density $\braket{n}_{\mathrm{GGE}}(\tau)$ as a function of $\tau=\Gamma_{\alpha}t$ from Eqs.~\eqref{eq:ann_interferences_dgl} and \eqref{eq:gfunction_interf}. (Right) Plot of the associated effective exponent $\delta_{\mathrm{eff}}(\tau)$ as a function of $\tau$ (log scale on the horizontal axis only). We set $d=1$ in Eq.~\eqref{eq:annihilation} for the two-body annihilation reaction and we show various values of $\theta=\pi/3,\pi/4,\pi/5$. \textbf{a)} Dynamics from the $\ket{\mbox{BEC}}$ initial state \eqref{eq:bc}. For $\theta=\pi/4$, the density is constant in time (top-red curve). For this reason, the effective exponent $\delta_{\mathrm{eff}}(\tau)$ is not shown on the right plot for $\theta=\pi/4$. For $\theta \neq \pi/4$, the density follows the mean-field decay $\braket{n}_{\mathrm{GGE}}(\tau)\sim \tau^{-1}$. \textbf{b)} Dynamics from the flat filling occupation function \eqref{eq:ff} for the same values of $\theta$ as in the previous panel. The density follows the mean-field decay for $\theta \neq 0,\pi/2$, while it obeys the slower non-mean-field decay $\braket{n}_{\mathrm{GGE}}(\tau)\sim \tau^{-1/2}$ for $\theta=\pi/4$. The right plot clearly shows the different asymptotic value of $\delta_{\mathrm{eff}}(\tau)\simeq 1/2$ for $\theta=\pi/4$ compared to the value $\delta_{\mathrm{eff}}(\tau)\simeq 1$ for $\theta \neq \pi/4$. \textbf{c)} Dynamics from the initial state \eqref{eq:gd} with Gaussian initial occupation function $B_q(0)$ for the same values of $\theta$ as in the previous panels. As in the case of panel (\textbf{b}), the non-mean-field decay is observed only for $\theta=\pi/4$.}
    \label{fig:interferences}
\end{figure}
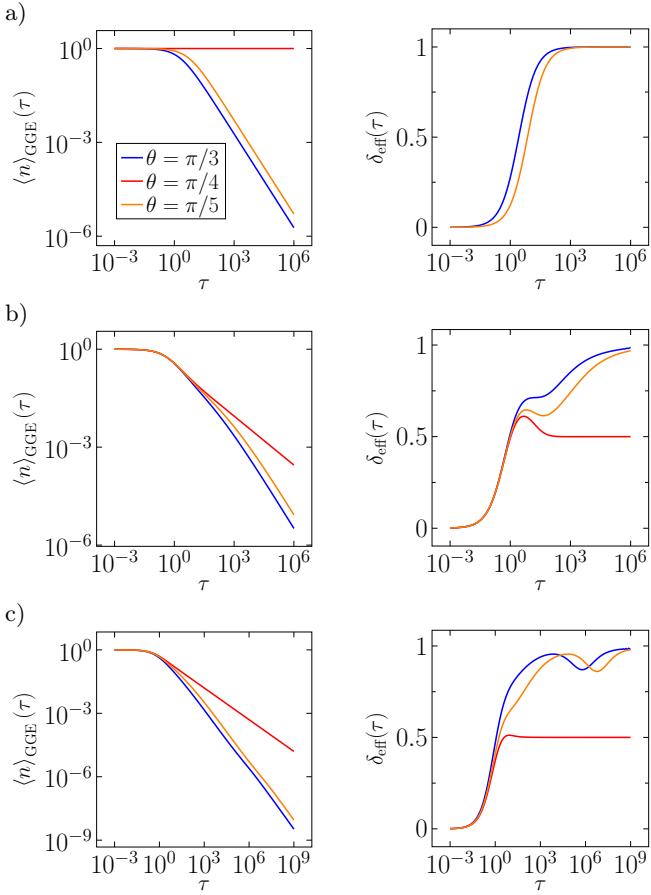

The TGGE rate equation \eqref{eq:qrd_diff} for the jump operators \eqref{eq:annihilation}, i.e. generic values of $\theta$, reads as (see Appendix \ref{app:binary_ann} for the details of the calculations)
\begin{equation} \label{eq:ann_interferences_dgl}
    \frac{d B_{q} (\tau)}{d\tau} = -\frac{1}{L} \sum_{k} g_{\theta, d} (k ,q) B_{k}(\tau) B_{q}(\tau),
\end{equation}
with $\tau = t \Gamma_{\alpha}$ the re-scaled time and 
\begin{align}
\notag
    g_{\theta, d} (k,q) = &2(1 + \cos(d(k -q))) - \sin(2  \theta) \\
    &\left[2 \cos(d(k+q)) + \cos(2 k d) + \cos(2qd) \right].
\label{eq:gfunction_interf}    
\end{align}
This differential equation is similar to the corresponding one for fermions of Ref.~\cite{QRD20222} (cf. Eq.~(S10) therein), the only difference being in the opposite sign of the terms $\cos(d(k-q))$ and $\cos(d(k+q))$. In Fig.~\ref{fig:interferences}, we plot the solution of the differential Eq.~\eqref{eq:ann_interferences_dgl} for the density $\braket{n}_{\mathrm{GGE}}(\tau)$ as a function of $\tau$. We find that the long-time dynamics at $\theta = \frac{\pi}{4}$ is qualitatively different than that obtained for all other values, i.e., $\theta \neq \pi/4$. Namely, for all the considered initial states, we find that for $\theta \neq \pi/4$, the density decay exponent is gain the mean-field one $\braket{n}_{\mathrm{GGE}}(\tau)\sim \tau^{-1}$. 

For $\theta=\pi/4$, however, the density is constant in time $\braket{n}_{\mathrm{GGE}}(\tau)=\braket{n}_{\mathrm{GGE}}(0)$. The BEC state is, indeed, a dark state of the jump operator in Eq.~\eqref{eq:annihilation} at $\theta = \frac{\pi}{4}$, which amounts to saying that $\ket{\mbox{BEC}}$ is annihilated by all the annihilation \eqref{eq:annihilation} jump operators
\begin{equation}
L_j^{\alpha}\ket{\mbox{BEC}}=0, \quad \forall j, \quad \theta=\pi/4.
\label{eq:dark_state}
\end{equation}
Moreover, the BEC is an eigenstate (the ground state in this case) of the Hamiltonian \eqref{eq:qrd_hamiltonian}. This observation is consistent with the results of Refs.~\cite{diehl2008,tomadin2011}, where the $\ket{\mbox{BEC}}$ state was similarly identified as an exact dark state of the quantum master equation of the noninteracting Bose gas. However, in Refs. \cite{diehl2008,tomadin2011}, the jump operators do not involve particle losses, but still they annihilate the antisymmetric part of the wavefunction due to terms $\hat{b}_i-\hat{b}_{i+d}$, thus rendering the BEC state dark. For the flat filling \eqref{eq:ff} and the Gaussian occupation states \eqref{eq:gd}, instead, we find a slower decay $\braket{n}_{\mathrm{GGE}}(\tau)\sim \tau^{-1/2}$ compared to mean field, as shown in Fig.~\ref{fig:interferences}. The asymptotic $1/2$ decay exponent (and the amplitude) can be exactly computed from the asymptotic of Eq.~\eqref{eq:ann_interferences_dgl} (see again Appendix \ref{app:binary_ann} for the details of the calculations). In particular, Eq.~\eqref{eq:ann_interferences_dgl} admits the following implicit solution for the mode occupation function $B_q(\tau)$
\begin{equation}
B_q(\tau)= B_q(0)\sqrt{\frac{n(\tau)}{n(0)}}\mbox{exp}\left[-2\sin^2(dq)\int_{0}^{\tau}\mbox{d}t \, n(t)\right].    
\label{eq:implicit_interference_1}
\end{equation}
Here, for the sake of brevity, we denoted $\braket{n}_{\mathrm{GGE}}(\tau)=n(\tau)$. The previous equation shows that $q=0$ is the slowest decaying mode. Furthermore, for $B_q(0)=n(0)\delta(q)$ (the BEC occupation in the limit $L\to \infty$), Eq.~\eqref{eq:implicit_interference_1} gives $n(\tau)=n(0)$ consistently with the previous discussion concerning the dark-state condition for the BEC. Using Eq.~\eqref{eq:implicit_interference_1} into Eq.~\eqref{eq:density_integral}, the $q$ integral for the density $n(\tau)$ can be asymptotically computed using the saddle-point approximation (which is justified since $\int_{0}^{\tau} \mbox{d}t\, n(t)\to \infty$ as $\tau \to \infty$). The density asymptotic one obtains 
\begin{equation}
n(\tau) \sim \left(\frac{B_{\pi}(0)+B_{0}(0)+2\sum_{n=1}^{d-1}B_{q^{\ast}_n}(0) }{4 d \sqrt{\pi n_0}}\right) \tau^{-1/2},
\label{eq:asymptotic_theta_pi4}
\end{equation}
with the saddle-points $q^{\ast}_n=n\pi/d$, $n=1,2\dots d-1$, determined by the $d-1$ zeros of the $\sin^2(dq)$ function in the interval $q\in(0,\pi)$. Therefore, when $\theta=\pi/4$ the decay exponent, $1/2$, does not depend on the distance $d$ between the pair of bosons involved in the loss. The amplitude of the decay, instead, does generically depend on $d$. Only in the case of the flat-filling initial occupation \eqref{eq:ff}, the amplitude of the decay takes the $d$-independent value $1/2 \sqrt{n_0/\pi}$. 

To sum up, for the decay exponent of the particle density in the presence of binary annihilation with interference between two decay channels \eqref{eq:annihilation}, we find
\begin{equation}
  \braket{n}_{\mathrm{GGE}}(\tau) \sim  \left\{ \begin{aligned} 
  &  \tau^{-1/2}  \quad &\mathrm{for} \quad \theta= \pi/4\\
  & \tau^{-1}  \quad &\mathrm{else}.  
\end{aligned} \right.,
\label{eq:result_theta_final}
\end{equation}
both for incoherent initial states \eqref{eq:ff} and for coherent ones, e.g., with a Gaussian initial occupation function $B_q(0)$ \eqref{eq:gd}. The same behavior \eqref{eq:result_theta_final} applies generically for initial quantum coherent states identified by a non-flat occupation function $B_q(0)$ of generic form. The $\ket{\mbox{BEC}}$ state \eqref{eq:bc} for $\theta=\pi/4$ does not decay since it is dark with respect to the dissipation. 

The result \eqref{eq:result_theta_final} significantly deviates from the one valid for fermions discussed in Ref.~\cite{QRD20222}. In the case of fermions, the decay $\braket{n}_{\mathrm{GGE}}(\tau)\sim \tau^{-1/2}$ is valid for any value of $\theta \neq 0,\pi/2$, not just $\theta=\pi/4$. The different behavior between fermions and bosons subject to binary annihilation with interference, and at the same time, the emergence of the beyond mean-field decay \eqref{eq:result_theta_final} can be understood by looking at the space continuum limit of Eq.~\eqref{eq:annihilation}. In the continuum limit, the lattice spacing $a$ (so far set to $1$) is reintroduced, so that lattice point are identified as $x_j=j a$ and $\ell=La$ is the dimensionful length of the system. The continuum limit is obtained by sending $a\to 0$, $L\to\infty$ with $\ell$ fixed, as we detail in Appendix \ref{app:cont_limit}. The jump operator \eqref{eq:annihilation} $L_{ja}^{\alpha}$ reduces, in this limit, to onsite pair annihilation $L_{ja}^{\alpha}\to L^{\alpha}(x) \sim \hat{b}^2(x)$ as long as $\theta\neq \pi/4$, with $\hat{b}(x)=\hat{b}_{ja}/\sqrt{a}$ being the continuum bosonic destruction field operator. This operator yields mean-field decay \eqref{eq:MF_decay_result_annihilation}, following the same steps as those done in Sec.~\ref{subsec:distance_selective_loss} for the lattice case. Only for $\theta=\pi/4$ (see again Appendix \ref{app:cont_limit} for the details), the binary annihilation operator \eqref{eq:annihilation} in the continuum limit takes a different form $L^{\alpha}(x) \sim \hat{b}(x)\partial_x \hat{b}(x)$. The latter explicitly couples to spatial derivatives of the bosonic field. The operator $L^{\alpha}(x)$ for $\theta=\pi/4$ therefore introduces spatial fluctuations in the quantum trajectories unravelling of the dynamics since reactions can more likely take place in depletion regions, where the density experiences spatial variations. This effect eventually causes the decay law \eqref{eq:result_theta_final}, which is beyond mean-field. In the fermionic case, the leading term $L^{\alpha}(x) \sim \hat{c}^2(x)$ of the continuum limit expansion, $\hat{c}(x)$ being the fermionic field destruction operator, is always zero because of the fermionic statistics. The continuum limit of fermionic nearest-neighbour annihilation is therefore always of the form $L^{\alpha}(x)\sim \hat{c}(x)\partial_x \hat{c}(x)$ and the decay $\braket{n}_{\mathrm{GGE}}(\tau)\sim \tau^{-1/2}$ applies for any $\theta \neq 0,\pi/2$. 

In order to further understand the emergence of non-mean-field decay exponents due to interfering decay channels, we proceed by discussing in the following Subsection annihilation reactions \eqref{eq:second_annihilation} with interferences among three different decay channels.

\subsection{Second-order annihilation} \label{subsec:second_order_annihilation}
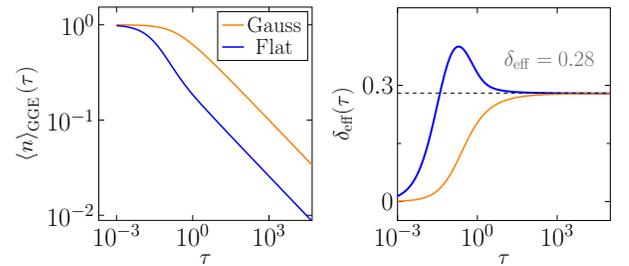
\begin{figure}[H]
    \centering
    \resizebox{0.474\columnwidth}{!}{\tikzsetnextfilename{second_order}
\begin{tikzpicture}
\begin{axis}[
            xmode=log,
            ymode=log,
            xmax=10^4,
            ylabel={$\braket{n}_{\mathrm{GGE}}(\tau)$},
            xlabel={$\tau$},
            enlargelimits=true,
            xtick={10^-3,10^0,10^3},
            ytick={10^-2,10^-1,10^0},
            yminorticks=false,
            ]
        \addplot[color=orange,  smooth,very thick,  scatter src=explicit symbolic] table[col sep=comma,header=true,x index=1,y index=2] {Data/sec_order_both_2.csv}; \addlegendentry{Gauss}
        \addplot[color=blue, smooth,very thick,  scatter src=explicit symbolic] table[col sep=comma,header=true,x index=1,y index=3] {Data/sec_order_both_2.csv}; \addlegendentry{Flat}
\end{axis}
\end{tikzpicture}}
    \resizebox{0.446\columnwidth}{!}{\tikzsetnextfilename{second_order_exponent}
\begin{tikzpicture}
\begin{axis}[
            xmode=log,
            ymax=0.5,
            xmin=10^-3,
            xmax=10^5,
            ylabel={$\delta_{\mathrm{eff}} (\tau)$},
            xlabel={$\tau$},
            minor y tick num=1,
            xtick={10^-3,10^0,10^3},
            ytick={0,0.3},
            ]
        \addplot[color=blue,smooth,ultra thick, scatter src=explicit symbolic] table[col sep=comma,header=true,x index=1,y index=3] {Data/sec_order_exp_both_2.csv};
        \addplot[color=orange, smooth, very thick,  scatter src=explicit symbolic] table[col sep=comma,header=true,x index=1,y index=2] {Data/sec_order_exp_both_2.csv};
        \draw[ycomb, dashed] (1e-3, 0.28) -- (1e9,0.28) node[black!60!white, below] at (5*1e2,0.4) {\LARGE $\delta_{\mathrm{eff}} = 0.28$}; 
\end{axis}
\end{tikzpicture}}
    \caption{\textbf{Quantum RD dynamics of second-order annihilation reaction.} Log-log plot (left) of the density $\braket{n}_{\mathrm{GGE}}(\tau)$ as a function of $\tau=\Gamma_{\bar{\alpha}}t$ from Eq.~\eqref{eq:sec_order_dgl}. In the right plot (log scale on the horizontal axis only), the effective exponent $\delta_{\mathrm{eff}}(\tau)$ as a function of $\tau$  (Eq.~\eqref{eq:eff_exponent}) is shown. Both the initial state \eqref{eq:ff} with a flat in $q$ occupation function, and \eqref{eq:gd} with an initial occupation function of Gaussian form are studied. 
    In both cases, the same algebraic decay $\braket{n}_{\mathrm{GGE}} (\tau) \sim \tau^{-0.28}$ is observed. The exponent of the algebraic decay is quantified by plotting the effective exponent, which at long times converges to $\delta_{\mathrm{eff}}(\tau)\simeq 0.28$ for both initial states \eqref{eq:ff} and \eqref{eq:gd}.}
    \label{fig:second_order}
\end{figure}

The TGGE rate equation \eqref{eq:qrd_diff} can be specialized to the jump operator \eqref{eq:second_annihilation} following similar steps as those performed in Secs.~\ref{subsec:distance_selective_loss} and \ref{subsec:annihilation_inteferences}. We report the main steps in Appendix \ref{app:binary_ann}, while here we give the final result:
\begin{equation} \label{eq:sec_order_dgl}
    \frac{\mbox{d} B_{q}(\tau)}{\mbox{d}\tau} = - \frac{4}{L} B_{q} \sum_{k}B_{k} \left[\cos(kd) + \cos(qd) - 2 \right]^{2},
\end{equation}
with the re-scaled time $\tau=\Gamma_{\bar{\alpha}}t$. The results of the numerical solution of the previous equation are reported in Fig.~\ref{fig:second_order}. The $\mbox{BEC}$ initial state \eqref{eq:bc} is also dark, as in Eq.~\eqref{eq:dark_state}, with respect to the jump operator $L_j^{\bar{\alpha}}$ (for any $j$) and therefore it is a many-body dark state of the quantum master equation. The density $\braket{n}_{\mathrm{GGE}}(\tau)=\braket{n}_{\mathrm{GGE}}(0)=n_0$ consequently remains constant in time. 
For both the flat filling \eqref{eq:ff} and the Gaussian occupation function \eqref{eq:gd} initial states, we, instead, observe algebraic decay asymptotically in time with a non-mean-field exponent 
\begin{equation}
    \braket{n}_{\mathrm{GGE}}(\tau) \sim \tau^{-0.28}.
\label{eq:interference_second_order_decay_results}
\end{equation}
This decay exponent is numerically computed by plotting the effective exponent $\delta_{\mathrm{eff}}(\tau)$ as a function of $\tau$, as shown in the right panel of Fig.~\ref{fig:second_order}. The effective exponent \eqref{eq:eff_exponent} converges at long times to the value $\delta_{\mathrm{eff}}(\tau)\simeq 0.28$ for both the initial states considered. The choice of the initial state solely affects how fast the asymptotic value of $\delta_{\mathrm{eff}}$ is met, but not the value itself. We also verified that the result in Eq.~\eqref{eq:interference_second_order_decay_results} holds generically for initial states identified by a different $q$-dependent initial occupation function $B_q(0)$. In the case of Eq.~\eqref{eq:interference_second_order_decay_results}, we remark, however, that it is not possible to derive analytically the asymptotic exponent, as we did in Eqs.~\eqref{eq:implicit_interference_1} and \eqref{eq:asymptotic_theta_pi4} of Subsec.~\ref{subsec:annihilation_inteferences}, due to the more complicated structure of Eq.~\eqref{eq:sec_order_dgl}.

The result \eqref{eq:interference_second_order_decay_results} is a non-mean-field prediction for the algebraic decay. Following the same reasoning used at the end of the previous Subsec.~\ref{subsec:annihilation_inteferences}, we can understand this behavior by looking at the continuum limit of the annihilation process \eqref{eq:second_annihilation} coupling three different decay channels. In this case, the continuum limit (see again Appendix \ref{app:cont_limit} for the details) leads to a jump operator of the form $L^{\bar{\alpha}}(x) \sim \hat{b}(x)\partial^2_x \hat{b}(x)$. The jump operator $L^{\bar{\alpha}}(x)$ couples now to second-order spatial derivatives (this is why we also name the process as ``second-order annihilation'') and it induces spatial fluctuations over larger regions and therefore  deviations from the mean field are more sensible. In particular, the exponent decreases compared to Eq.~\eqref{eq:asymptotic_theta_pi4} for the first-order interference drifting further from the mean-field value ($\delta=1$). 

Apparently, interference effects generally slow down the decay of the density compared to the classical reaction-limited case. In both cases analysed in Eqs.~\eqref{eq:asymptotic_theta_pi4} and \eqref{eq:interference_second_order_decay_results}, we see that quantum interferences are necessary in order to observe beyond mean-field decay in the noninteracting Bose gas. 

\subsection{Coagulation} \label{subsec:coagulation}
\begin{figure}[t]
    \centering
    \subfloat{\resizebox{0.48\columnwidth}{!}{\tikzsetnextfilename{coagulation}
\begin{tikzpicture}
\begin{axis}[
            xmode=log,
            ymode=log,
            ylabel={$\braket{n}_{\mathrm{GGE}}(\tau)$},
            xlabel={$\tau$},
            enlargelimits=true,
            ]
        \addplot[color=red,smooth,thick, scatter src=explicit symbolic] table[col sep=comma,header=true,x index=1,y index=2] {Data/coagulation_density.csv};
\end{axis}
\end{tikzpicture}}}
    \hfill
     \subfloat{\resizebox{0.45\columnwidth}{!}{\tikzsetnextfilename{coagulation_exponent}
\begin{tikzpicture}
\begin{axis}[
            xmode=log,
            ylabel={$\delta_{\text{eff}}(\tau)$},
            xlabel={$\tau$},
            enlargelimits=true,
            ytick={1, 0.5,0},
            minor y tick num=1,
            ]
        \addplot[color=red,smooth,thick, scatter src=explicit symbolic] table[col sep=comma,header=true,x index=1,y index=2] {Data/coagulation_delta.csv};
\end{axis}
\end{tikzpicture}  }}
    \caption{\textbf{Coagulation quantum RD dynamics.} Log-log plot (left) of the density $\braket{n}_{\mathrm{GGE}}(\tau)$ as a function of time $\tau$. The initial condition in the given case is the flat filling state in Eq.~\eqref{eq:ff}, but the very same curve is obtained for the BEC \eqref{eq:bc} and the Gaussian occupation state \eqref{eq:gd}. In all the cases, the density decays as in the mean-field approximation $\braket{n}_{\mathrm{GGE}} \sim \tau^{-1}$. In the right plot, the effective exponent $\delta_{\mathrm{eff}}(\tau)$ \eqref{eq:eff_exponent} is plotted (log scale on the horizontal axis only) and it converges to mean-field value $1$ at long times.}
    \label{fig:coagulation_dynamics}
\end{figure}
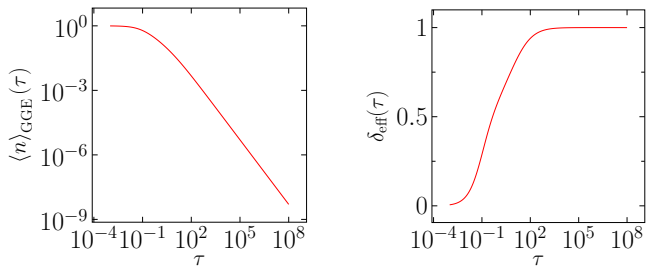

The onsite coagulation decay is modelled by the jump operator \eqref{eq:coagulation} introduced in Sec.~\ref{sec:system}. It describes the reaction $2A\to A$ where two (or more) bosons meet on the same site $j$ leading to the destruction of one of the particles at the same lattice site. The resulting rate equation (cf. Appendix \ref{app:coagulation} for the details of the calculations) for the bosonic occupation function is
\begin{equation}
    \frac{\mbox{d} B_q(\tau)}{\mbox{d}\tau} = -6 B_q \braket{n}_{\mathrm{GGE}}^2 + 2 \braket{n}_{\mathrm{GGE}}^2 -4 B_q \braket{n}_{\mathrm{GGE}},
\label{eq:coagulation_B_q}
\end{equation}
with $\tau=\Gamma_{\gamma}t$ the re-scaled time according to the coagulation rate. We note that for one-site coagulation, similarly to the case of onsite binary annihilation \eqref{eq:MF_onsite_results}, the TGGE equation for the density is closed, and it reads  
\begin{equation}
\label{eq:coag_dgl}
    \frac{\mbox{d} \braket{n}_{\mathrm{GGE}}}{\mbox{d}\tau} =-\left(6 \braket{n}_{\mathrm{GGE}}^{3} + 2 \braket{n}_{\mathrm{GGE}}^{2}  \right).
\end{equation}
This equation contains both a two-body term, $\braket{n}_{\mathrm{GGE}}^2$, as in the corresponding classical law of mass action \eqref{eq:rd_rl}, and a three-body term, $\braket{n}_{\mathrm{GGE}}^3$. The latter is notably absent in the classical law of mass action description. It is also absent in the fermionic quantum reaction-limited dynamics \cite{QRD20222} and therefore a genuine property of the quantum bosonic reaction-limited dynamics. In the presence of coagulation only (and more generally in the presence of reactions depleting the system) this three-body term does not, however, impact the late time asymptotics. The density $\braket{n}_{\mathrm{GGE}}$, as a matter of fact, decays asymptotically in time to zero and therefore the three-body term $\braket{n}_{\mathrm{GGE}}^{3}$ can be neglected with respect to the two-body term $\braket{n}_{\mathrm{GGE}}^{2}$. The latter immediately gives the mean-field algebraic decay
\begin{equation}
    \braket{n}_{\mathrm{GGE}}(\tau) \sim \tau^{-1}.
\label{eq:coagulation_MF_result}
\end{equation}
In Fig.~\ref{fig:coagulation_dynamics}, we report the numerical solution of Eq.~\eqref{eq:coagulation_B_q} together with the plot of the associated effective exponent $\delta_{\mathrm{eff}}(\tau)$ versus the re-scaled time $\tau$. The effective exponent converges to the mean-field exponent $\delta_{\mathrm{eff}}(\tau)\simeq 1$ at long times. This mean-field behaviour is, importantly, valid for any initial state \eqref{eq:bc}-\eqref{eq:gd}, quantum coherent or not. A similar result has been derived in Ref.~\cite{QRD20222} for fermionic systems, where the coagulation decay similarly shows $\braket{n}_{\mathrm{GGE}}(\tau= \Gamma_{\gamma}t) = \tau^{-1}$ decay both for coherent and incoherent initial states. In the fermionic case, for a Fermi-sea initial state, quantum coherences are only observed in Ref.~\cite{QRD20222} to rescale time by a factor dependent on the initial density $n_0$, leaving the asymptotic decay exponent unchanged. In the bosonic case, instead, Eq.~\eqref{eq:coag_dgl} gives the very same dynamics for all kind of initial conditions. We also note that the coefficient $2$ in front of the two-body term on the right-hand side of Eq.~\eqref{eq:coag_dgl} is half of the coefficient $4$ of the onsite annihilation two-body term \eqref{eq:MF_onsite_results}. This implies that the density $\braket{n}_{\mathrm{GGE}}^{\mathrm{coag/ann}}(\tau,n_0)$ ($n_0$ being the initial density) as a function of time in the annihilation and coagulation processes are asymptotically (as long as the $\braket{n}_{\mathrm{GGE}}^3$ term for coagulation can be neglected) related by 
\begin{equation}
\braket{n}_{\mathrm{GGE}}^{\mathrm{coag}}(\tau,n_0)=2 \braket{n}_{\mathrm{GGE}}^{\mathrm{ann}}(\tau,n_0/2), \,\,\, \mbox{for}\,\,\, \Gamma_{\alpha}=\Gamma_{\gamma}.
\label{eq:annihilation_coagulation_equiv}
\end{equation}
In the context of classical RD  \cite{henkel1995equivalences,henkel1997reaction,krebs1995finite,simon1995concentration,ben2005relation} this relation is taken as the hallmark that signals that both coagulation and annihilation dynamics belong to the same universality class. In particular, this implies that the two processes share the same asymptotic decay. Apparently in the bosonic quantum RD system the two processes still belong to the same universality class, at least in the reaction-limited regime. This statement applies to both quantum coherent and incoherent initial states, in contrast with the fermionic case of Ref.~\cite{QRD20222}, where the equivalence \eqref{eq:annihilation_coagulation_equiv} applies only for incoherent initial states.   

\subsection{Absorbing-state phase transition in quantum bosonic RD} \label{subsec:branching}
In order to obtain a process with a stationary state featuring a non-zero density of particles, we need to include reactions increasing the particle number, thereby competing with the loss processes \eqref{eq:annihilation}-\eqref{eq:coagulation}. Therefore, we consider onsite branching reactions $A\to A+A$ \eqref{eq:branching} at rate $\Gamma_{\beta}$. For the branching reaction process, we found the following rate equation (see Appendix \ref{app:branching_bosons} for the calculation):
\begin{equation}
\frac{\mbox{d}B_q(t)}{\mbox{d}t}=\Gamma_{\beta}[ 2(2n-B_q)+8n^2 +6 B_q n^2].
\label{eq:branching_mode_occupation_onsite}
\end{equation}
In the following, we denote $\braket{n}_{\mathrm{GGE}}(t)=n$, dropping the time dependence for the sake of brevity. As in the case of onsite coagulation \eqref{eq:coag_dgl}, the particle density obeys the closed equation
\begin{equation}
\frac{\mbox{d}n(t)}{\mbox{d}t}=\Gamma_{\beta}[2n +8n^2 +6n^3].
\label{eq:MF_result_onsite_br}
\end{equation}
We note that, as well as for coagulation, branching produces the three-body term, $n^3$, which is absent both in the classical and in the fermionic reaction-limited RD dynamics. The branching process leads to the creation of particles and therefore  the density increases in time in Eq.~\eqref{eq:MF_result_onsite_br}.

This persistent growth can be countered by introducing annihilation reactions, and in the following we will study an absorbing-state phase transition that results from such competing processes.  
We consequently consider a model where (onsite) branching \eqref{eq:branching}, at rate $\Gamma_{\beta}=\Gamma \beta$, is competing with (onsite) coagulation \eqref{eq:coagulation}, at rate $\Gamma_{\gamma}=\Gamma \gamma$, and one-body decay \eqref{eq:1b_decay}, at rate $\Gamma_{\delta}=\Gamma \delta$. This parametrization of the rates allows to identify $\Gamma$ as the overall dissipation (inverse) time scale, while $\beta,\gamma,\delta$ encode the relative strength of the three considered processes. Considering Eq.~\eqref{eq:MF_result_onsite_br} together with \eqref{eq:coag_dgl} one then finds the rate equation
\begin{align}
   \frac{\mbox{d} n}{\mbox{d}\tau} = (2\beta-\delta)n +(8\beta-2\gamma)n^2 +6(\beta-\gamma)n^3,
\label{eq:CP_bosons_MF_results_all}
\end{align}
where $\tau=\Gamma t$. Note, that the linear term, $-\delta n$, stems from the one-body decay process. 

We already note that Eq.~\eqref{eq:CP_bosons_MF_results_all} is qualitatively different from the classical mean-field description \eqref{eq:MF_contact_process} of the unrestricted contact process due to the presence of cubic and quartic terms in the density. However, Eq.~\eqref{eq:CP_bosons_MF_results_all} is still of mean-field type since it implies that the GGE is factorized also in real space: $\rho_{\mathrm{GGE}}(t)\sim \mbox{exp}(-\beta(t) \hat{N})$. In this sense, for bosons, the quantum mean-field description is different and richer than the classical mean-field one.
Eq.~\eqref{eq:CP_bosons_MF_results_all} can be equivalently written as 
\begin{equation}
\frac{\mbox{d}n(\tau)}{\mbox{d}\tau}=-\frac{\partial W(n)}{\partial n},    
\label{eq:quantum_CP_potential_0}
\end{equation}
where  
\begin{equation}
W(n)=(\delta-2\beta)\frac{n^2}{2}+(2\gamma-8\beta)\frac{n^3}{3}+ 6(\gamma-\beta)\frac{n^4}{4},
\label{eq:quantum_CP_potential}
\end{equation}
has the meaning of a potential. We remark that a similar fourth order potential $W(n)$ was found in Refs.~\cite{marcuzzi2016,buchhold2017}, which studied the so-called quantum contact process. Microscopically this process is, however, different to the one discussed here: it is formulated in terms of spin $1/2$ particles (not bosons) and the Hamiltonian embodies coherent branching and coagulation, i.e., spin flips conditional on the excitation of a neighbouring spin. In our case, instead, the Hamiltonian \eqref{eq:qrd_hamiltonian} leads to free, unconstrained, hopping of particles. The dissipation is described by classical incoherent processes in both cases. Despite their apparent differences both models yield the same potential $W(n)$. 

It is noteworthy that in Refs.~\cite{marcuzzi2016,buchhold2017} the potential $W(n)$ describes the stationary phases of the quantum contact process only in the absence of spatial fluctuations, i.e., within the mean-field approximation. The latter is valid only in high spatial dimensions, while it fails in one dimension. In the present work, however, $W(n)$ emerges in one dimension in the reaction limited regime $\Gamma/\Omega \ll 1$ as an exact result. 

\begin{figure}[t]
    \centering
    \includegraphics[width=1\columnwidth]{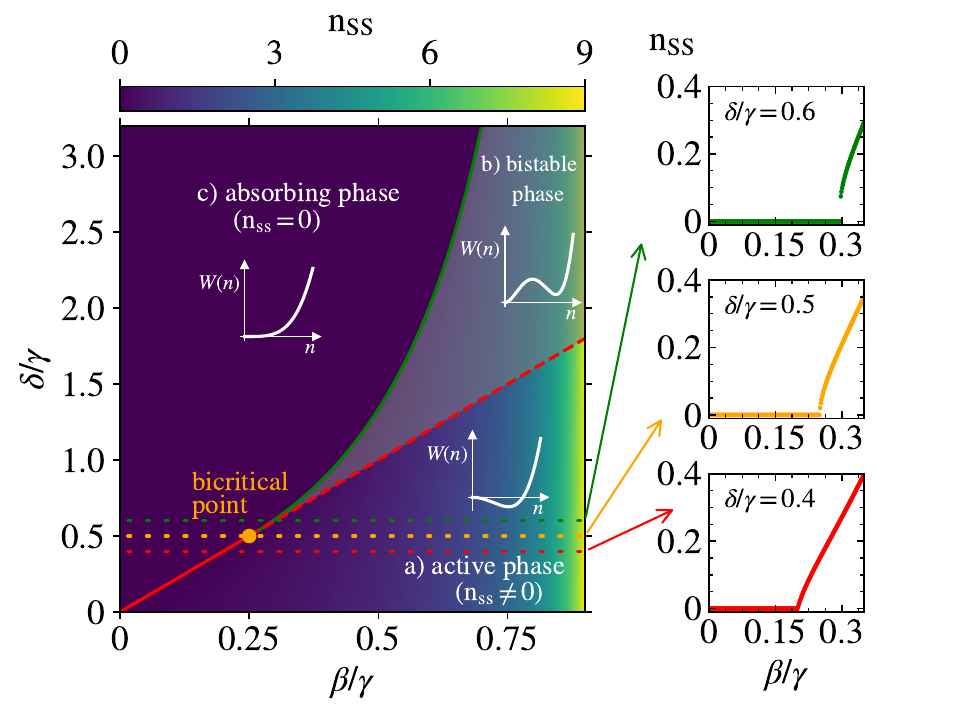}
    \caption{\textbf{Stationary phase diagram of quantum RD with incoherent branching and coagulation}. The stationary state density $n_{\mathrm{SS}}$, represented by the color code, is plotted as a function of the ratios $\beta/\gamma$, incoherent branching over incoherent coagulation, and $\delta/\gamma$, incoherent decay over incoherent coagulation. 
    We identify three different regions $\mathrm{a)-c)}$ in the phase diagram. In each region, we sketch the potential $W(n)$, Eq. (\ref{eq:quantum_CP_potential}). A second-order continuous phase transition separates the absorbing $\mathrm{c})$ and the active $\mathrm{a})$ phases and it is plotted with the red-solid line $\beta=\delta/2$. 
    The stationary density $n_{\mathrm{SS}}$ vanishes continuously as a function of $\beta/\gamma$, as shown in the bottom-right panel for a fixed value of $\delta/\gamma=0.4$. This transition belongs to the mean-field classical directed percolation universality class \eqref{eq:2nd_order_classical_transition_beta}.  A first-order discontinuous phase transition takes place across the green-solid line \eqref{eq:delta_critical}. This line separates the absorbing phase $\mathrm{c})$ from a bistable phase $\mathrm{b})$ (shown in grey-blurred scale). The bistable phase is delimited by the coexistence lines $(\delta/\gamma)_c$ (green solid) and $\delta=2\beta$ (red dashed). Across the first-order line, the density discontinuously jumps from zero to the value $n_{\mathrm{SS}}$, as shown in the top-right panel for a fixed value of $\delta/\gamma=0.6$ (green-dotted horizontal cut). The two coexistence lines meet at the bicritical point (orange dot) $(\beta/\gamma, \delta/\gamma) =(1/4, 1/2)$. At this point, the transition is still second order, as shown in the central-right panel, but it belongs to the mean-field tricritical directed percolation universality class \eqref{eq:bicritical_1} and \eqref{eq:bicritical_2}.}
    \label{fig:phase_diagram}
\end{figure}

The stationary-state phase diagram of the process \eqref{eq:quantum_CP_potential_0} is reported in Fig.~\ref{fig:phase_diagram}, where we plot the stationary state density $n_{\mathrm{SS}}$ (represent by the color code) as a function of the ratios $\beta/\gamma$ and $\delta/\gamma$. The stationary density $n_{\mathrm{SS}}$ is one of the (possibly multiple) minima of $W(n)$. The number of minima of $W(n)$ and their position depends on $\beta/\gamma$ and $\delta/\gamma$. For $\beta/\gamma>1$, the potential $W(n)$ is unbounded from below and the active phase supports an infinite density. For this reason, we take $\beta/\gamma<1$, in the following. In the Fig.~\ref{fig:phase_diagram}, we also sketch the potential $W(n)$ so that the stationary state density $n_{\mathrm{SS}}$ value in the various regions of the phase diagram can be readily understood. We can identify three different phases $\mathrm{a})-\mathrm{c})$:
\begin{itemize}
\item[a)] If $\delta/\gamma < 2\beta/\gamma$, an active, i.e. finite density solution, $n_{\mathrm{act}}\neq 0$,  of Eq.~\eqref{eq:CP_bosons_MF_results_all} is present:
\begin{equation}
n_{\mathrm{act}}= \frac{8\beta-2\gamma+\sqrt{(8\beta-2\gamma)^2-24(\gamma-\beta)(\delta-2\beta)}}{12(\gamma-\beta)}.
\label{eq:stationary_active_solution}
\end{equation}
The density $n_{\mathrm{act}}$ is the only minimum of $W(n)$, while $n_{\mathrm{abs}}=0$ is a maximum. In this case, the system is therefore in the \textit{active phase} $n_{\mathrm{SS}}=n_{\mathrm{act}}$.
\item[b)] For $\beta/\gamma>1/4$ we define the curve $(\delta/\gamma)_c$ as
\begin{equation}
\left(\frac{\delta}{\gamma}\right)_c= \frac{2\beta}{\gamma} +\frac{1}{24}\frac{(2-8\beta/\gamma)^2}{1-\beta/\gamma}.
\label{eq:delta_critical}
\end{equation}
For $\delta/\gamma<(\delta/\gamma)_c$ and $\delta/\gamma>2\beta/\gamma$, the potential $W(n)$ has two minima $n_{\mathrm{abs}}=0$ and $n_{\mathrm{act}}$ \eqref{eq:stationary_active_solution}. For this choice of parameters the stationary state $n_{\mathrm{SS}}$ can be either $n_{\mathrm{abs}}=0$ or $n_{\mathrm{act}}$ depending on the value of the initial density $n(\tau=0)=n_0$. If $n_0$ is small enough so as it falls in the basin of attraction of $n_{\mathrm{abs}}=0$, then $n_{\mathrm{SS}}=0$. On the contrary, if $n_0$ is large enough so as to fall in the basin of attraction of $n_{\mathrm{act}}$, then the system is active $n_{\mathrm{SS}}=n_{\mathrm{act}}$. For this reason, we identify this region of the phase diagram as a \textit{bistable phase}. 
\item[c)] For $\beta/\gamma>1/4$ and $\delta/\gamma > (\delta/\gamma)_c$, as well as for $\delta/\gamma>2\beta/\gamma$ and $\beta/\gamma<1/4$, the potential $W(n)$ has a single minimum $n_{\mathrm{abs}}=0$. The system is therefore in the \textit{absorbing phase} $n_{\mathrm{SS}}=0$.
\end{itemize}

For $\beta/\gamma<1/4$, the boundary between a) and c) is given by the line $\beta/\delta=1/2$, which is shown in red solid in Fig.~\ref{fig:phase_diagram}. Along this line the order parameter $n_{\mathrm{SS}}$ of the transition varies continuously. This is shown in the bottom-right panel of Fig.~\ref{fig:phase_diagram}, where the stationary density $n_{\mathrm{SS}}$ is displayed as a function $\beta/\gamma$ for $\delta/\gamma=0.4$ (horizontal red-dotted line in the density plot in Fig.~\ref{fig:phase_diagram}). In particular, expanding Eq.~\eqref{eq:stationary_active_solution} for $\beta-2\delta\to 0$ and $\beta< \gamma/4$ one obtains
\begin{equation}
n_{\mathrm{SS}} \sim (\beta/\delta-1/2)^{1}, \quad \beta/\gamma<1/4.
\label{eq:2nd_order_classical_transition_beta}
\end{equation}
At the same time, setting $\beta/\delta=1/2$ and $\beta/\gamma<1/4$, one finds that the density $n(t)$ decays asymptotically as
\begin{equation}
n(t)\sim t^{-1}, \quad \beta/\delta=1/2, \quad \beta/\gamma<1/4.
\label{eq:2nd_order_classical_transition_delta}
\end{equation}
Equations \eqref{eq:2nd_order_classical_transition_beta} and \eqref{eq:2nd_order_classical_transition_delta} identify the critical exponents of the classical mean-field directed percolation universality class \cite{vladimir1997nonequilibrium,henkel2008non,hinrichsen2000non,tauber2014critical,Malte_Henkel_2004}, as also recalled after Eq.~\eqref{eq:MF_CP_bosons}. Along the second-order transition line, the cubic nonlinearity $n^3$ of \eqref{eq:CP_bosons_MF_results_all} is therefore negligible and the transition is driven by the second-order $n^2$ nonlinearity. This renders \eqref{eq:MF_CP_bosons} effectively analogous to the classical mean-field equation \eqref{eq:MF_CP_bosons}. The classical mean-field directed percolation exponents are therefore obtained. 
For $\beta/\gamma>1/4$, the boundary between b) and c) is given by the curve $(\delta/\gamma)_c$ in Eq.~\eqref{eq:delta_critical} and it reveals, instead, a richer physics compared to the mean-field description \eqref{eq:MF_contact_process} of the classical contact process. 
In particular, along the line $(\delta/\gamma)_c$ given by \eqref{eq:delta_critical}, plotted in solid green in Fig.~\ref{fig:phase_diagram}, we find a first-order transition since the density jumps discontinuously from the value \eqref{eq:stationary_active_solution} to zero. This is shown in the middle panel in Fig.~\ref{fig:phase_diagram}. Therein we plot the stationary density $n_{\mathrm{SS}}$ as a function of $\beta/\gamma$ for fixed $\delta/\gamma=0.6$ (horizontal green-dotted line). 

Equation \eqref{eq:delta_critical}, for the first-order line, is obtained by calculating when the potential $W(n)$ has an inflection point $\mbox{d}^2W(n)/\mbox{d}n^2=0$. For $\delta/\gamma <(\delta/\gamma)_c$ and $\beta/\gamma >1/4$, the inflection point morphs into a first minimum $n_{\mathrm{act}}$ \eqref{eq:stationary_active_solution}, while a second minimum is given by $n_{\mathrm{abs}}=0$. We emphasize that the relative height of the two minima $W(n_{\mathrm{act}})/W(n_{\mathrm{abs}})$ does not fix the stationary density since in this nonequilibrium case there is no principle of free-energy minimization as in equilibrium systems. In the latter case, indeed, $W(n)$ would be interpreted as a free energy and the equilibrium solution would necessarily be the absolute minimum of $W(n)$ in the thermodynamic limit. In the present nonequilibrium scenario, instead, the stationary density can take either the value $n_{\mathrm{SS}}=n_{\mathrm{act}}$ or the value $n_{\mathrm{SS}}=n_{\mathrm{abs}}=0$ depending on the initial density $n_0$. 
This phenomenon is referred to as bistability and it happens in proximity of the first-order branch $(\delta/\gamma)_c$, where the absorbing and the active phase coexist due to the presence of two minima in $W(n)$. The bistable phase (plotted in grey blurred scale in Fig.~\ref{fig:phase_diagram}) is delimited by the coexistence curves $(\delta/\gamma)_c$ (green-solid line) and $\delta=2\beta$ (red-dashed line).  Coexistence of difference phases in the presence of a first-order phase transition has been also observed in a different context, both in equilibrium and in nonequilibrium, for Dicke-Bose-Hubbard systems \cite{landig2016quantum,bistability}.

The two coexistence lines meet the second-order line at the bicritical point $(\beta/\gamma,\delta/\gamma)=(1/4,1/2)$ (represented in orange in Fig.~\ref{fig:phase_diagram}). At this point, the transition is continuous, but it belongs to a different universality class than that of Eqs.~\eqref{eq:2nd_order_classical_transition_beta} and \eqref{eq:2nd_order_classical_transition_delta}. Namely, for $\beta/\gamma=1/4$ and $2\beta-\delta \to 0$, we find from \eqref{eq:stationary_active_solution} the power law 
\begin{equation}
n_{\mathrm{SS}} \sim (\beta/\delta -1/2)^{1/2}, \quad \beta/\gamma=1/4,    
\label{eq:bicritical_1}
\end{equation}
for the stationary density as a function of $\beta/\delta$. This is shown in the central-right panel of Fig.~\ref{fig:phase_diagram}, where $n_{\mathrm{SS}}$ is plotted as a function of $\beta/\gamma$ for fixed $\delta/\gamma=0.5$ (horizontal orange-dashed line in the density plot in Fig.~\ref{fig:phase_diagram}). The  jump discontinuity in $n_{\mathrm{SS}}$ across the first-order branch goes continuously to zero as the bicritical point is approached. Similarly, at $\beta/\delta=1/2$ and $\beta=\gamma=1/4$, one has that the density $n(t)$ decays algebraically in time as
\begin{equation}
n(t) \sim t^{-1/2}, \quad \beta/\delta=1/2, \quad \beta/\gamma=1/4.
\label{eq:bicritical_2}
\end{equation}
The same critical exponents as in Eqs.~\eqref{eq:bicritical_1} and \eqref{eq:bicritical_2} have been found also in Refs.~\cite{marcuzzi2016,buchhold2017} for the mean-field regime of the quantum contact process involving both classical-incoherent and quantum-coherent branching/coagulation. Here, a similar bicritical point to that of Fig.~\ref{fig:phase_diagram} was found in the plane spanned by the classical and quantum branching rates (note the difference with Fig.~\ref{fig:phase_diagram}, where both the axes $\beta/\gamma$ and $\delta/\gamma$ are classical rates). The bicritical point is argued in Refs.~\cite{marcuzzi2016,buchhold2017} to be described by the tricritical directed percolation universality class \cite{grassberger1982,lubeck2006tricritical,grassberger2006tricritical}. For spatial dimension $D\geq 3$, the mean-field exponents \eqref{eq:bicritical_1} and \eqref{eq:bicritical_2} are found, while in $D<3$ deviations from mean field are present. The results of the present manuscript suggest that the same universality class describes also the bicritical point of the model involving coherent hopping \eqref{eq:qrd_hamiltonian} and classical branching/coagulation. The mean-field values for the exponents \eqref{eq:bicritical_1} and \eqref{eq:bicritical_2} are then a consequence of considering the reaction-limited regime $\Gamma/\Omega \ll 1$, where spatial fluctuations can be neglected. Consequently, in the reaction-limited regime here considered, such exponents are exact already in one spatial dimension $D=1$.

\section{Discussion} \label{sec:discussion}
In this manuscript, we have studied the quantum reaction-limited reaction-diffusion dynamics of the noninteracting Bose gas. The main finding of the manuscript is that bosonic statistics considerably impacts on the universal properties of the quantum RD dynamics. In particular, we find that the combination of bosonic statistics with quantum effects, such as coherent hopping and quantum interferences, leads to universal nonequilibrium behavior different from both fermionic and classical RD systems.

The dynamics for QRD systems is formulated in terms of the quantum master equation \eqref{eq:lindblad_master}-\eqref{eq:qrd_dissipator}.
Within this formulation, we considered various types of reactions introduced in Sec.~\ref{sec:system}. 
Our analytical study is based on the time-dependent generalized Gibbs ensemble method, briefly recalled in Sec.~\ref{sec:rd_dynamics}. This method gives direct access to the quantum reaction-diffusion dynamics in the thermodynamic limit in the so-called reaction-limited regime $\Gamma/\Omega\ll1$ of weak dissipation. 
\begin{table}[t]
    \centering
    \begin{tabular}{l|c|c|c}
     $L_{j}^{\nu}$  & BEC &  flat filling & Gaussian \\ \hline
   $\sqrt{\Gamma_{\alpha}} \hat{b}_{j}\hat{b}_{j+d}$ & 1 & 1 & 1 \\
   \hline
   $\sqrt{\Gamma_{\alpha}/2} \hat{b}_{j} \left( \hat{b}_{j+d} - \hat{b}_{j-d}\right)$  & 0 & 1/2 & 1/2 \\
   \hline
   $\sqrt{\Gamma_{\Bar{\alpha}}} \hat{b}_{j} \left( \hat{b}_{j+1} + \hat{b}_{j-1} - 2 \hat{b}_{j}\right)$  & 0 & 0.28 & 0.28 \\
   \hline
   $\sqrt{\Gamma_{\gamma}} \hat{n}_{j}\hat{b}_{j}$  & 1 & 1 & 1 \\
   \hline  
    \end{tabular}
    \caption{\textbf{Summary of quantum reaction-limited density decay exponents for the noninteracting Bose gas.} In this table, we summarise our results from Eq.~\eqref{eq:tGGE} for the different annihilation reaction types \eqref{eq:annihilation}-\eqref{eq:coagulation} treated in Subsecs.~\ref{subsec:distance_selective_loss}-\ref{subsec:coagulation} in the reaction-limited regime $\Gamma/\Omega \ll 1$. In all the listed cases, the density decays as a power-law $\braket{n}_{\mathrm{GGE}}(\tau) = \tau^{-\delta}$ with the listed exponent $\delta$. The exponent is given for each reaction and for the considered initial states. The flat filling \eqref{eq:ff} is incoherent, while the BEC \eqref{eq:bc} and the Gaussian state \eqref{eq:gd} are quantum coherent. In the case of the BEC, the value $0$ of the exponent, on the second and third line, refers to the fact that the BEC is dark to these jump operators and therefore it does not decay.}
    \label{tab:summary_exponents}
\end{table}
Our results for the density algebraic decay $\braket{n}_{\mathrm{GGE}}(\tau)\sim \tau^{-\delta}$ in the presence of only reactions depleting the system are summarized in Table \ref{tab:summary_exponents}. We see that the law of mass action \eqref{eq:rl_crd} decay exponent is generically recovered for all classical-incoherent reactions, such as distance selective losses \eqref{eq:annihilation} ($\theta \neq 0$) and coagulation. 
This result applies both to quantum coherent and incoherent initial states. 
Algebraic decay beyond mean field is possible for bosons only when interference between different decay channels is allowed, as shown in Figs.~\ref{fig:interferences} and \ref{fig:second_order}. In the case of first-order interference \eqref{eq:annihilation} with $\theta=\pi/4$, in Subsec.~\ref{subsec:annihilation_inteferences}, we find, namely, a power-law decay with exponent $1/2$ (cf. Eq.~\eqref{eq:asymptotic_theta_pi4}). For all the other values $\theta \neq \pi/4$, mean-field decay is observed. For the second-order interference \eqref{eq:second_annihilation} in Subsec.~\ref{subsec:second_order_annihilation}, the decay exponent is, instead, approximately $0.28$. Incoherent coagulation \eqref{eq:coagulation} decay also yields mean-field decay \eqref{eq:coagulation_MF_result}. 
These results show markedly different behavior between bosons and fermions. In the latter case, power-law decay beyond mean field is, indeed, attained under broader conditions, such as also for classical annihilation reactions from incoherent initial states, and for generic interference decay $\theta \neq 0,\pi/2$ \cite{QRD20222,perfetto2023quantum,riggio2023effects}. Furthermore, for bosons, coagulation and binary annihilation belong to the same universality class, at least as long as the reaction-limited regime is concerned. For fermions, instead, this is not true \cite{QRD20222}.

In table \ref{tab:summary_exponents_2}, we summarize the main critical exponents associated to the absorbing-state phase transition observed in bosonic systems in the presence of branching \eqref{eq:branching}, coagulation and decay.  
\begin{table}[H]
    \centering
    \begin{tabular}{p{3cm}|p{1.5cm}|p{3.5cm}}
     contact process \newline $(\beta, \gamma, \delta)$& 2nd order \newline transition \newline $\beta=\delta/2$ &  bicritical point \newline $(\beta/\gamma, \delta/\gamma) = (1/4,1/2)$ \\ \hline
     $n_{\mathrm{SS}} \sim (\beta/\delta-1/2)^{\xi}$ & 1& 1/2 \\ \hline
     $n(t)\sim t^{-\chi}$ & 1 &1/2 \\ \hline
    \end{tabular}
    \caption{\textbf{Summary of quantum reaction-limited exponents for the absorbing-state phase transition in the noninteracting Bose gas.} In this table, we summarise our results for the quantum RD model with absorbing-state phase transition in the reaction-limited regime (see Subsec.~\ref{subsec:branching}). This transition can be either of first or second order. In the table, we summarize the exponents $(\xi,\chi)$ associated to the latter. In the second column, the exponents associated to the second-order line $\beta=\delta/2$ are reported. These exponents are those of the mean-field directed percolation universality class. We further give (third column) the exponents associated with the bicritical point of Fig.~\ref{fig:phase_diagram}. These exponents are those of the mean-field tricritical directed percolation universality class.}
    \label{tab:summary_exponents_2}
\end{table}
The corresponding stationary-state phase diagram in Fig.~\ref{fig:phase_diagram} is of mean-field nature. However, it is qualitatively different from both the classical mean-field \eqref{eq:MF_contact_process} and the fermionic analogue \cite{QRD20222}.
In the bosonic case, we find that both a first-order and a second-order line are present.The first-order transition line has no classical counterpart. The second-order line is again characterized by the critical exponents of the mean-field directed percolation universality class (middle column of Table \ref{tab:summary_exponents_2}) as in the classical reaction-limited description \eqref{eq:MF_contact_process}. Interestingly, the second-order line and the first-order one meet at a bicritical point. This point is identified by the mean-field exponents of the tricritical directed percolation universality class (rightmost column of Table \ref{tab:summary_exponents_2}). The universality class of the bicritical point is consequently different from directed percolation, which characterizes the classical dynamics. For bosons, the quantum mean-field description of the contact process phase diagram is therefore richer than the classical mean-field description in Eq.~\eqref{eq:MF_contact_process}.

Our work paves the way for many future research directions concerning the Bose gas subject to dissipative loss processes. It would be interesting to study the quantum RD dynamics in the presence of reactions conserving the particle number as in Refs.~\cite{diehl2008,tomadin2011}. Therein, dissipation allows for the dynamical preparation of a pure $\ket{\mbox{BEC}}$ state. This state is a many-body dark state of the dynamics and it is therefore attained at long times. It is important and experimentally relevant to study the stability of this dynamical preparation protocol of the BEC state against particle losses. 

It is also interesting to look at different observables than the particle density. The two-point bosonic correlation function $\braket{\hat{b}^{\dagger}_n \hat{b}_m}_{\mathrm{GGE}}(\tau)$ can be readily computed within the TGGE method and its temporal decay would shed light on the decay of quantum coherences as the stationary state is approached. In addition, it is interesting to study Bose superfluids with weak dissipative losses. These models can be also addressed with the TGGE ansatz of this manuscript. We expect that for Bose superfluids the dynamics of the superfluid order parameter gives additional information on the decoherence in time of the system due to dissipation, as it has been shown in Ref.~\cite{lossth11} for lossy fermionic superfluids.

Ultimately, it is desirable to go beyond the quantum reaction-limited regime $\Gamma/\Omega\ll 1$ discussed in this manuscript. The quantum analogue of the diffusion-limited decay \eqref{eq:diffusion_limited_intro} is, indeed, not known.
Quantifying the anticipated quantum diffusion-limited algebraic decay might be possible by exploiting the field theory formulation of the quantum master equation via the Keldysh path integral method \cite{kamenev2023field, sieberer2016keldysh}. In the diffusion-limited regime, the significance of spatial fluctuations due to the finite hopping is inherently increased. A systematic renormalization group scaling analysis is therefore needed in order to quantify the space dependence and the impact of spatial fluctuations on the decay exponents. 

With respect to the absorbing-state phase transition, it is important to understand the reason behind the similarity between the phase diagram of Fig.~\ref{fig:phase_diagram} and the one found for the quantum contact process in Refs.~\cite{marcuzzi2016,buchhold2017}. The model studied therein is different from the quantum RD model here proposed since coherent effects are introduced via a different Hamiltonian giving coherent branching/coagulation. In our case the Hamiltonian \eqref{eq:qrd_hamiltonian} simply gives free hopping. The field-theory scaling analysis of the quantum RD model of Subsec.~\ref{subsec:branching} could, in this way, shed light on the, largely not understood, universality class of the quantum contact process away from the reaction-limited, mean-field, limit. 
\vspace{-0.35cm}
\acknowledgments
\vspace{-0.35cm}
We are grateful for financing from the Baden-W\"urttemberg Stiftung through Project No.~BWST\_ISF2019-23. We also acknowledge funding from the Deutsche Forschungsgemeinschaft (DFG, German Research Foundation) under Project No. 435696605, through the Research Unit FOR 5413/1, Grant No. 465199066 and through the Research Unit FOR 5522/1, Grant No. 499180199. G.P.~acknowledges support from the Alexander von Humboldt Foundation through a Humboldt research fellowship for postdoctoral researchers. 

\onecolumngrid

\appendix
\section{Continuum Limit of two-body annihilation with interferences}  \label{app:cont_limit}
In this Appendix, we derive the space continuum limit of the annihilation operators \eqref{eq:annihilation} and \eqref{eq:second_annihilation}. This continuum limit provides an argument for the emergence of non-mean-field algebraic decay exponents [see Eqs. \eqref{eq:asymptotic_theta_pi4} and \eqref{eq:interference_second_order_decay_results}] in the noninteracting Bose gas subject to binary annihilation processes. 

In order to perform the continuum limit, we reintroduce the lattice spacing $a$ (formerly set to 1), which provides the shortest length scale of the problem. The dimensionful length of the chain is $\ell=La$, where $L$ is the number of lattice sites. Lattice points are identified as $x_j=ja$ in units of the lattice spacing. The space continuum limit is achieved by taking the limit $a\to 0$, $L\to \infty$ with $\ell$ kept fixed. In this limit, the Lindblad equation \eqref{eq:lindblad_master}-\eqref{eq:qrd_dissipator} can be written as
\begin{equation}
\frac{\mbox{d}\rho}{\mbox{d}t}=-i[H,\rho]+\sum_{\nu}\int_{0}^{\ell} \mbox{d}x \left[L^{\nu}(x)\rho L^{\nu \dagger}(x)-\frac{1}{2}\left\{L^{\nu \dagger}(x)L^{\nu}(x),\rho\right\} \right].
\label{eq:Lindblad_space_continuum}
\end{equation}
In the previous equation, the Hamiltonian $H$ is obtained from the continuum limit of Eq. \eqref{eq:qrd_hamiltonian} as
\begin{equation}
H=-\Omega \sum_{j=1}^{L} \left( \hat{b}_{ja}^{\dagger}\hat{b}_{ja+a} + \hat{b}_{ja+a}^{\dagger}\hat{b}_{a} \right) \quad \rightarrow \quad H=\int_{0}^{\ell} \mbox{d}x \, \hat{b}^{\dagger}(x) (-\overline{\Omega} \partial^2_{x})\hat{b}(x), 
\label{eq:qrd_hamiltonian_continuum}
\end{equation}
with $\overline{\Omega}=\Omega a^2$. The continuum bosonic field operators $\hat{b}(x)$ and the jump operators $L^{\nu}(x)$  are defined as
\begin{equation}
\hat{b}(x)=\frac{\hat{b}_{ja}}{\sqrt{a}}, \quad \mbox{with} \quad  [\hat{b}(x),\hat{b}^{\dagger}(x')]=\delta(x-x'),\quad \mbox{and} \quad L^{\nu}(x)=\frac{L_{ja}^{\nu}}{\sqrt{a}}.
\end{equation}
The Hamiltonian \eqref{eq:qrd_hamiltonian_continuum} is the second-quantized Hamiltonian of free bosonic particles freely moving in free space, with the kinetic energy operator $\sim \partial_x^2$ proportional to the second partial derivative in the space coordinate $x$. The definition of the jump operators $L^{\nu}(x)$ includes a factor $\sqrt{a}$, which is needed in order to turn the sum over lattice sites in Eq. \eqref{eq:qrd_dissipator} into a space integral as in Eq. \eqref{eq:Lindblad_space_continuum}. In the remainder of this and the other Appendices, we denote bosonic operators $\hat{b}_j, \hat{b}_j^{\dagger}$ (and their continuum analogues $\hat{b}(x)$, $\hat{b}^{\dagger}(x)$) as $b_j, b_j^{\dagger}$, without the hat symbol for the sake of brevity. We use the same notation also for bosonic number operators $\hat{n}_j, \hat{n}_q$ in real and momentum space, respectively, which we denote as $n_j, n_q$.   

The continuum limit $L^{\alpha}(x)$ of the binary annihilation jump operator can be obtained from the Taylor series of Eq.~\eqref{eq:annihilation}   
\begin{align}
 L^{\alpha}(x)=L_{ja}^{\nu}/\sqrt{a}&=\sqrt{\Gamma_{\alpha}/a} \, b_{ja} \left( \cos(\theta) b_{ja+da} - \sin(\theta) b_{ja + da}  \right) \nonumber \\
 &=\sqrt{\Gamma_{\alpha}^{(1)}}b^2(x)+\sqrt{\Gamma_{\alpha}^{(2)}}b(x)\partial_x b(x)+\mathcal{O}(a^{5/2}), 
 \label{eq:continuum_1st_interf}
\end{align}
where the coefficients $\Gamma_{\alpha}^{(1/2)}$ are given by 
\begin{equation}
\Gamma_{\alpha}^{(1)}=a\Gamma_{\alpha}(\cos (\theta)-\sin(\theta))^2, \quad \mbox{and} \quad \Gamma_{\alpha}^{(2)}= a^{3} d^2 \Gamma_{\alpha} \, (\cos(\theta)+\sin(\theta))^2. 
\label{eq:continuum_rates_1st_interf}
\end{equation}
We can see that $\Gamma_{\alpha}^{(1)}$ is the leading term as $a\to 0$, 
implying that $L^{\alpha}(x)\sim b^2(x)$ reduces to a binary annihilation process. The point $\theta=\pi/4$ is, instead, special because in that case one has $\Gamma_{\alpha}^{(1)}=0$ and the next term $\Gamma_{\alpha}^{(2)}$ becomes the leading one in the expansion. Such term produces the coupling $L^{\alpha}(x)\sim b(x)\partial_x b(x)$ between the annihilation process and spatial derivatives of the bosonic field. This property introduces spatial fluctuations in the dynamics, since the action of the jump operator is sensitive to spatial gradients of the density profile. This yields eventually yields the beyond mean-field decay law \eqref{eq:asymptotic_theta_pi4}.

For the second-order annihilation process \eqref{eq:second_annihilation}, it is immediate to verify that Eq. \eqref{eq:second_annihilation} in the continuum limit yields a second order derivative of the bosonic field $b(x)$. Namely, one has
\begin{align}
L^{\Bar{\alpha}}(x)=L^{\Bar{\alpha}}_{ja}/\sqrt{a}&=\sqrt{\Gamma_{\Bar{\alpha}}/a}b_{ja}(b_{ja+da}+b_{ja-da}-2b_{ja}) \nonumber \\
&=\sqrt{\Gamma_{\Bar{\alpha}}^{(2)}}b(x)\partial_x^2 b(x),
\label{eq:continuum_2nd_interf}
\end{align}
with the coefficient $\Gamma_{\Bar{\alpha}}^{(2)}=a^5 d^4 \Gamma^{\Bar{\alpha}}$. The jump operator \eqref{eq:continuum_2nd_interf} couples to second-order spatial derivatives of the bosonic field and therefore it accounts for spatial fluctuations of larger spatial extent. This causes the beyond mean-field decay of Eq. \eqref{eq:interference_second_order_decay_results}. In particular, the second space derivative probes larger space regions than the first one and therefore the decay exponent \eqref{eq:interference_second_order_decay_results} decreases compared to that of Eq.~\eqref{eq:asymptotic_theta_pi4}. On the basis of this reasoning, we expect the decay exponent to become smaller the higher the order of the interference process (higher order spatial derivatives).

\section{Two-Body annihilation dynamics} \label{app:binary_ann}
In this Appendix, we briefly report the main steps of the derivation of Eqs.~\eqref{eq:ann_interferences_dgl} and \eqref{eq:sec_order_dgl} for the binary annihilation process ($2A \to \emptyset$) involving interferences of two \eqref{eq:annihilation} or three \eqref{eq:second_annihilation} decay channels, respectively. We consider throughout the case of periodic boundary conditions $b_{j+L} = b_{j}$ for the Hamiltonian in Eq.~\eqref{eq:qrd_hamiltonian}. As this derivation is based on the TGGE method we are taking the thermodynamic limit $L \to \infty$, the choice of the periodic boundary condition does not therefore change the final result. The Fourier transform of the bosonic operators $b_{j}, b_{j}^{\dagger}$ is defined as follows
\begin{equation} \label{eq:fourier}
    b_{k_{n}} = \frac{1}{\sqrt{L}} \sum_{j=1}^{L} e^{-i k_{n} j} b_{n}, \quad \text{with inverse} \quad b_{j} =   \frac{1}{\sqrt{L}} \sum_{k_{n}=\frac{2 \pi}{L}}^{2\pi} e^{i k_{n} j} b_{k_{n}},
\end{equation}
where $k_{n} = 2\pi n/L$, with the integer number $n =1,2 \dots, L$, are the quasi-momenta. The Fourier transformed operators satisfy the bosonic commutation relations $[b_k,b_{k'}^{\dagger}]=\delta_{k,k'}$. In the following, we write the quasi-momenta $k_{n}$ as $k$ for brevity. Sums $\sum_{k_n}=\sum_{n=1}^{L}$ over the integer $n=1,2\dots L$ will be denoted as $\sum_{k}$ for simplicity. When sum over multiple quasi-momenta $k_{1} , \dots , k_{L}$ are present, e.g., $\sum_{k_{1}} \dots \sum_{k_{N}}$, we use the shorter notation $\sum_{k_{1},\dots,k_{N}}$. In the thermodynamic limit $L\to \infty$, when the TGGE applies, the momenta become continuous variables $k\in (-\pi,\pi)$ and the summations over $k_n$ turn to integrals as in Eq.~\eqref{eq:density_integral}. 

The Fourier transform of the jump operator $L_j^{\alpha}$ [Eq. \eqref{eq:annihilation}] for the binary annihilation reaction gives,
\begin{equation}
\label{eq:fourier_jump_first_order}
    L_{j}^{\alpha} = \frac{\sqrt{\Gamma_{\alpha}}}{L} \sum_{k,k'} b_{k}b_{k'} e^{ij(k+k')} \left( \cos(\theta) e^{idk'} - \sin(\theta) e^{-idk'} \right).
\end{equation}
To explicitly write the right hand side of the rate equation in Eq.~\eqref{eq:qrd_diff}, we need to evaluate the commutator $[n_q,L_j^{\alpha}]$. For this purpose, the following commutation relations are useful
\begin{equation} \label{eq:commutator_relation}
    \left[n_{q} , b_{k}b_{k'} \right] = - b_{k}  b_{k'} \left(  \delta_{kq} + \delta_{k'q}\right), \quad \mathrm{from} \quad [n_{q}, b_{k}] = -\delta_{kq} b_{q}.
\end{equation}
Plugging the above relation \eqref{eq:commutator_relation} into the rate equation \eqref{eq:qrd_diff}, we find
\begin{equation}
     \frac{\mbox{d} B_{q}}{\mbox{d}t} = - \frac{\Gamma_\alpha}{L} \sum_{k_{1}, k_{2}, k,k'}  \delta_{k+k',k_1+k_2} f^{\ast}_{\theta,d}(k_1)f_{\theta,d}(k')
     \left(\braket{b_{k_1}^{\dagger}b_{k_2}^{\dagger}b_{k}b_{q}}_{\mathrm{GGE}}\delta_{q,k'}+\braket{b_{k_1}^{\dagger}b_{k_2}^{\dagger}b_{k'}b_{q}}_{\mathrm{GGE}}\delta_{q,k}\right),
\label{eq:1st_step_annihilation}
\end{equation}
where we used the Fourier representation of the Kronecker delta 
\begin{equation}
\delta_{k,k'}\equiv \delta_{k_n,k_m}=\frac{1}{L}\sum_{j=1}^{L}e^{i j 2\pi (n-m)/L}.
\label{eq:kronecker_delta}
\end{equation}
In Eq.~\eqref{eq:1st_step_annihilation}, we further defined the function 
\begin{equation}
f_{\theta,d}(k)=\cos(\theta)e^{ik d}-\sin(\theta)e^{-ikd}. \label{eq:f_function_annihilation}   
\end{equation}
The four-point bosonic correlation functions in Eq.~\eqref{eq:1st_step_annihilation} can be crucially decomposed in terms of the two-point bosonic occupation function $\braket{b_q^{\dagger}b_k}=B_q \delta_{q,k}$ since the GGE state \eqref{eq:tGGE} is Gaussian. This is achieved via application of Wick's theorem
\begin{equation}
\braket{b^{\dagger}_{k_1} b^{\dagger}_{k_2} b_{k_1+k_2-q}b_q}=B_{k_1}B_q \delta_{k_2,q}+B_{k_1}B_{k_2}\delta_{k_1,q}.
\label{eq:Wick_annihilation}
\end{equation}
Using Eqs.~\eqref{eq:kronecker_delta}-\eqref{eq:Wick_annihilation} to evaluate Eq.~\eqref{eq:1st_step_annihilation} one obtains
\begin{equation} \label{eq:ann_interferences_dgl_app}
    \frac{\mbox{d} B_{q} (\tau)}{\mbox{d}\tau} = -\frac{1}{L} \sum_{k} g_{\theta, d} (k ,q) B_{k}(\tau) B_{q}(\tau),
\end{equation}
with $\tau=\Gamma_{\alpha}t$ the rescaled time. The function $g_{\theta,d}(k,q)$ is obtained from \eqref{eq:f_function_annihilation} as
\begin{align}
    g_{\theta, d} (k,q) &= 2\mbox{Re}(f^{\ast}_{\theta,d}(k) f_{\theta,d}(q))+ |f_{\theta,d}(q)|^2+|f_{\theta,d}(k)|^2 \nonumber \\
    &= 2(1 + \cos(d(k -q))) - \sin(2  \theta) \left[2 \cos(d(k+q)) + \cos(2 k d) + \cos(2q d) \right].
\label{eq:gfunction_annihilation_supp}
\end{align}
Equations \eqref{eq:ann_interferences_dgl_app} and \eqref{eq:gfunction_annihilation_supp} coincide with Eqs.~\eqref{eq:ann_interferences_dgl} and \eqref{eq:gfunction_interf} of the main text. Upon setting $\theta=0$, Eq.~\eqref{eq:gfunction_annihilation_supp} immediately renders \eqref{eq:sol_distance_selective}. For $\theta=\pi/4$, the expression in Eq.~\eqref{eq:gfunction_annihilation_supp} simplifies to \begin{equation}
g_{\pi/4,d}(k,q)= 2[\sin(kd) +\sin(qd)]^2=2[\sin^2(kd)+\sin^2(qd)+2\sin(kd)\sin(qd)].
\label{eq:gfunction_pi4}
\end{equation}
This function is very similar to the one reported in Refs.~\cite{lossth1,QRD20222} for the case of fermions (setting $d=1$ therein). The only difference with respect to the fermionic case lies in the sign change $[\sin(kd)-\sin(qd)]^2$, which in turn causes the last term on the right hand side of the second equality of Eq. \eqref{eq:gfunction_pi4} to have the opposite sign $-2\sin(kd)\sin(qd)$. This difference, however, is not important as long as one considers initial states whose distribution is even in $q$: $B_{q}(0)=B_{-q}(0)$. This is true for all the initial states \eqref{eq:bc}-\eqref{eq:gd} we considered in this work. If the initial state is, indeed, invariant under quasi-momenta reversal $q\to-q$, then this symmetry is kept at all times $B_q(\tau)=B_{-q}(\tau)$. This, in turn, implies that the last term on the right hand side the second equality of \eqref{eq:gfunction_pi4} is zero since $\sum_{k}\sin(kd) B_k(\tau)=0$ (the $\sin(kd)$ function is odd in $k$). In this case, Eq.~\eqref{eq:ann_interferences_dgl} reduces to the very same form discussed in Refs.~\cite{lossth1,QRD20222} for fermions:
\begin{equation}
 \label{eq:ann_interferences_dgl_pi4}
    \frac{d B_{q} (\tau)}{d\tau} = -\frac{1}{L} \sum_{k} g_{\pi/4, d} (k ,q) B_{k}(\tau) B_{q}(\tau)=-\frac{1}{L} \sum_{k} [2\sin^2(kd)+2\sin^2(qd)] B_{k}(\tau) B_{q}(\tau).
\end{equation}    
From this equation, we can derive the implicit solution \eqref{eq:implicit_interference_1} for $B_q(\tau)$, following similar derivations as those performed in Refs.~\cite{lossth1,lossth2}. We report here this derivation for the sake of completeness. The equation for the density $\braket{n}_{\mathrm{GGE}}(\tau)\equiv n(t)$ from \eqref{eq:ann_interferences_dgl_app} reads as 
\begin{equation}
\frac{\mbox{d}n(\tau)}{\mbox{d}\tau}= -4 n(\tau)\frac{1}{L}\sum_{k}\sin^2(kd)B_k(\tau) \quad \rightarrow \quad \frac{1}{L}\sum_{k}\sin^2(kd)B_k(\tau)= -\frac{1}{4 n(\tau)}\frac{\mbox{d}n(\tau)}{\mbox{d}\tau}. 
\label{eq:intermediate_analytics_pi4}
\end{equation}
We remark that \eqref{eq:intermediate_analytics_pi4} is not a closed equation for the density $n(t)$ since the occupation function $B_k(\tau)$ appears explicitly on the right hand side. The evolution for the occupation function $B_q$ of the mode $q$ is, indeed, coupled to that of all the other modes $B_k$ through the function $g_{\pi/4,d}(k,q)$. This function arises from the structure of the jump operator \eqref{eq:annihilation} involving the superposition of two decay channels between nearest neighbouring sites. In the case of onsite processes, e.g., coagulation and branching discussed in Appendix \ref{app:coagulation} and \ref{app:branching_bosons} below, respectively, different Fourier modes are not coupled and one can write closed equations for the density of particles. Substituting Eq.~\eqref{eq:intermediate_analytics_pi4} into \eqref{eq:ann_interferences_dgl_pi4} yields
\begin{equation}
\frac{d B_{q} (\tau)}{d\tau}=\frac{B_q \dot{n}(\tau)}{2 n(\tau)}-2\sin^2(dq) B_q(\tau)n(\tau),    
\label{eq:second_intermediate_analytics}
\end{equation}
with $\dot{n}(\tau)=\mbox{d}n(\tau)/\mbox{d}\tau$. The time integration of Eq.~\eqref{eq:second_intermediate_analytics} gives
\begin{equation}
\ln\left(B_q(\tau)/B_q(0)\right)=\frac{1}{2}\ln\left(n(t)/n(0)\right)-2\sin^2(qd)\int_{0}^{\tau}\mbox{d}t \, n(t).
\label{eq:analytics_final}
\end{equation}
The exponentiation of \eqref{eq:analytics_final} eventually yields Eq.~\eqref{eq:implicit_interference_1} of the main text.

The derivation of Eq.~\eqref{eq:sec_order_dgl} for the jump operator $L_j^{\Bar{\alpha}}$ \eqref{eq:second_annihilation} describing second-order annihilation decay follows the very same steps as those leading to \eqref{eq:ann_interferences_dgl_app}. In particular, Eq.~\eqref{eq:1st_step_annihilation} is still valid upon changing the definition of the function $f_{\theta,d}(k) \to f_{d}(k)$. For the jump operator \eqref{eq:second_annihilation} one has
\begin{equation}
f_d(k)=2(\cos(kd)-1).
\label{eq:2nd_annihilation_f_function}
\end{equation}
Then, Eq.~\eqref{eq:ann_interferences_dgl_app} is still valid with $g_{\theta,d}(k)\to g_{d}(k)$ and 
\begin{equation}
g_{d}(k)= 2\mbox{Re}(f^{\ast}_{d}(k) f_{d}(q))+ |f_{d}(q)|^2+|f_{d}(k)|^2=4(\cos(kd)+\cos(qd)-2)^2.
\label{eq:gfunction_2nd_order_app}
\end{equation}
Eqs.~\eqref{eq:ann_interferences_dgl_app} and \eqref{eq:gfunction_2nd_order_app} coincide with \eqref{eq:sec_order_dgl} of the main text. Both in the cases of Eq.~\eqref{eq:gfunction_annihilation_supp} and in that of Eq.~\eqref{eq:gfunction_2nd_order_app}, we solve numerically the TGGE rate equation \eqref{eq:ann_interferences_dgl_app} by discretizing the momenta $k_n=2\pi n/L$ in the interval $(-\pi,\pi)$. Here $L$ parametrizes the resolution of the grid. We take the value $L=10^4$. We have checked that the numerical data in Figs.~\ref{fig:ann_distances}-\ref{fig:second_order} are stable upon further increasing $L$.

\section{Coagulation dynamics}
\label{app:coagulation}
For the onsite coagulation dynamics in Eq.~\eqref{eq:coagulation}, the Fourier transform of the jump operator $L_j^{\gamma}$ reads 
\begin{equation} \label{eq:fourier_coag}
    L_{j}^{\gamma} = \sqrt{\frac{\Gamma_\gamma}{L^3}} \sum_{k_1, k_2, k_3} e^{ij(k_2 + k_3 - k_1)} b^{\dagger}_{k_1} b_{k_2} b_{k_3}.
\end{equation}
The commutator $[n_q,L_j^{\gamma}]$ in Eq.~\eqref{eq:master_tgge} yields
\begin{align} \label{eq:comm_coag}
    [n_q , b^{\dagger}_{k_1} b_{k_2}b_{k_3}] = -b^{\dagger}_{k_1}b_{k_2}b_q \delta_{q,k_3}-b^{\dagger}_{k_1}b_{k_3}b_q \delta_{q,k_2}+b^{\dagger}_q b_{k_2}b_{k_3}\delta_{q,k_1}.
\end{align}
Inserting \eqref{eq:fourier_coag} (and the adjoint equation for $L_j^{\gamma \dagger}$) into \eqref{eq:master_tgge} and using \eqref{eq:comm_coag} to handle the commutator, one obtains
\begin{align}
     \frac{\mbox{d} B_q (t)}{\mbox{d}t} &= \frac{\Gamma_\gamma}{L^{2}} \sum_{k,k',k_1,k_2}\braket{b^{\dagger}_k b^{\dagger}_{k'}b_{k+k'-k_1-k_2+q} b^{\dagger}_q b_{k_1}b_{k_2}}_{\mathrm{GGE}}-2 \braket{b^{\dagger}_k b^{\dagger}_{k'}b_{k+k'+k_1-k_2-q} b^{\dagger}_{k_1} b_{k_2}b_{q}}_{\mathrm{GGE}}.
\label{eq:coag_intermediate_app}
\end{align}
The six-point bosonic correlation functions can be again computed exploiting the Gaussian structure of the GGE and therefore Wick's theorem. One has
\begin{align}
\braket{b^{\dagger}_k b^{\dagger}_{k'}b_{k+k'-k_1-k_2+q}b^{\dagger}_q b_{k_1}b_{k_2}}_{\mathrm{GGE}}&= B_{k'}B_q B_k\delta_{k',k_2}\delta_{q,k_1}+B_{k'}B_q B_k \delta_{k',k_1}\delta_{q,k_2}+B_k B_q B_{k'}\delta_{k,k_1}\delta_{q,k_2}\nonumber \\
&+B_{k'}B_k B_q \delta_{k,k_2}\delta_{q,k_1}+B_k B_{k'}(1+B_q)\delta_{k,k_1}\delta_{k',k_2}+B_{k}B_{k'}(1+B_q)\delta_{k,k_2}\delta_{k',k_1}.
\label{eq:Wick_six_point_coag_1}
\end{align}
and 
\begin{align}
\braket{b^{\dagger}_k b^{\dagger}_{k'}b_{k+k'+k_1-k_2-q}b^{\dagger}_{k_1} b_{k_2}b_{q}}_{\mathrm{GGE}}&= B_{k'}B_{k_1} B_k\delta_{k',k_2}\delta_{q,k_1}+B_{k'}B_{k_1} B_k \delta_{k',q}\delta_{k_1,k_2}+B_k B_{k_1} B_{k'}\delta_{k,k_2}\delta_{q,k_1}\nonumber \\
&+B_{k'}B_k B_{k_1} \delta_{k,q}\delta_{k_1,k_2}+B_k B_{k'}(1+B_{k_1})\delta_{k,k_2}\delta_{k',q}+B_{k}B_{k'}(1+B_{k_1})\delta_{k',k_2}\delta_{k,q}.
\label{eq:Wick_six_point_coag_2}
\end{align}
Inserting \eqref{eq:Wick_six_point_coag_1} and \eqref{eq:Wick_six_point_coag_2} into \eqref{eq:coag_intermediate_app}, one eventually obtains
\begin{equation}
\frac{\mbox{d} B_q(\tau)}{\mbox{d}\tau} = -6 B_q \braket{n}_{\mathrm{GGE}}^2 + 2 \braket{n}_{\mathrm{GGE}}^2 -4 B_q \braket{n}_{\mathrm{GGE}},
\end{equation}
with $\tau=\Gamma_{\gamma}t$. This equation coincides with Eq.~\eqref{eq:coagulation_B_q} of the main text. In this equation, crucially, the occupation $B_q$ of the mode $q$ is not coupled to that of the other modes $k\neq q$. This allows to write the closed equation \eqref{eq:coag_dgl} for the density. Therein three-body terms $\sim \braket{n}_{\mathrm{GGE}}^3$ are present. Note, that in the analogous calculation for the fermions \cite{QRD20222} such terms cancel out, but here, for bosons, they do not. These cubic nonlinearities are the origin of the rich phase diagram emerging when coagulation competes with branching. We present the calculations for branching in the next Section of the Appendix.

\section{Branching dynamics}
\label{app:branching_bosons}
We report here the main steps of derivation the rate equation Eq.~\eqref{eq:branching_mode_occupation_onsite}. The steps are very similar to those of Appendix \ref{app:coagulation} and we therefore report only the main steps. The Fourier transform of the onsite branching jump operator $L_j^{\beta}$ in Eq.~\eqref{eq:branching} reads
\begin{equation} \label{eq:fourier_jump_branching}
    L_{j}^{\beta} = \sqrt{\frac{\Gamma_{\beta}}{L^{3}}} \sum_{k_{1}, k_{2}, k_{3}} b_{k_{1}}^{\dagger} b_{k_{2}}^{\dagger} b_{k_{3}} e^{ij(k_{3} - k_{1} - k_{2})}.
\end{equation}
The commutator $[n_q,L_j^{\beta}]$ in Eq.~\eqref{eq:fourier_jump_branching} is readily computed from the knowledge of the following commutators
 \begin{equation}
     [n_{q}, b_{k_{1}}^{\dagger} b_{k_{2}}^{\dagger} b_{k_{3}}] = b_{q}^{\dagger} b_{k_{2}}^{\dagger} b_{k_{3}} \delta_{k_{1},q} +  b_{q}^{\dagger} b_{k_{1}}^{\dagger} b_{k_{3}} \delta_{k_{2},q} - b_{k_{1}}^{\dagger} b_{k_{2}}^{\dagger} b_{q} \delta_{k_{3},q}.
\label{eq:commutators_branching}     
 \end{equation}
Inserting Eqs.~\eqref{eq:fourier_jump_branching} and \eqref{eq:commutators_branching} into \eqref{eq:master_tgge} one obtains
\begin{align} \label{eq:branching_dgl_wick}
    \frac{\mbox{d}B_{q}}{\mbox{d}t} = \frac{\Gamma_{\beta}}{L^{2}} &\sum_{k,k',k_{1},k_{2}}2\braket{b^{\dagger}_k b_{k'}b_{k-k'-k_2+k_1+q}b_q^{\dagger}b^{\dagger}_{k_1}b_{k_2}}_{\mathrm{GGE}}-\braket{b^{\dagger}_k b_{k'}b_{k-k'-q+k_2+k_1}b^{\dagger}_{k_1}b^{\dagger}_{k_2}b_q}_{\mathrm{GGE}}.
\end{align}
The calculation of the six-point functions follows to very same lines of the calculation of Eqs.~\eqref{eq:Wick_six_point_coag_1} and \eqref{eq:Wick_six_point_coag_2} and we therefore do not report it here for the sake of brevity. The final result for the evolution equation for the occupation function $B_q$ in momentum space reads
\begin{equation}
\frac{\mbox{d}B_q(t)}{\mbox{d}t}=\Gamma_{\beta}[ 2(2n-B_q)+8n^2 +6 B_q n^2],    
\end{equation}
which coincides with Eq.~\eqref{eq:branching_mode_occupation_onsite} of the main text. As in the case of onsite coagulation, for onsite branching the evolution of a mode $q$ is not coupled to the evolution of the other modes $k\neq q$. This allows to derive the closed equations \eqref{eq:MF_result_onsite_br} and \eqref{eq:CP_bosons_MF_results_all} for the density of particles. These equations are akin to the classical law of mass action equation \eqref{eq:MF_contact_process}, but, at the same time, they are fundamentally different from those equations due to the three body terms. These terms generate the phase diagram discussed in Subsec.~\ref{subsec:branching} of the main text, which shows a richer behavior than the classical mean-field contact process in Eq.~\eqref{eq:MF_contact_process}.

\twocolumngrid
\bibliography{bib}

\end{document}